\documentclass[final,onefignum,onetabnum]{siamart190516}
\usepackage{etex}
\usepackage{morewrites}
\usepackage{etoolbox}
\usepackage{xparse}
\usepackage[T1]{fontenc} 
\usepackage[utf8]{inputenc} 
\usepackage{amssymb,amsfonts,amsopn,dsfont,amscd}
\usepackage{latexsym}
\usepackage{bm}
\usepackage{mathtools}
\usepackage{nicefrac}
\usepackage{mathrsfs}
\usepackage{mathdots}
\usepackage{commath}
\usepackage{bm}
\usepackage{stmaryrd}
\usepackage{empheq}
\usepackage{cancel}
\usepackage{enumitem}
\usepackage{pifont}
\usepackage{tikz}
\usepackage{pgfplots}
\usepackage{tikz-cd}
\usepackage{longtable}
\usepackage{float}
\usepackage{fancybox}
\usepackage{tcolorbox}
\usepackage{colortbl}
\usepackage{listings}
\usepackage{lstautogobble}
\usepackage{subfig}
\usepackage{ifthen}
\usepackage{cite}
\usepackage{listings}
\lstdefinestyle{pseudocode}{
  basicstyle=\sffamily\color{black},    
  morecomment=[l]{//},                  
  morecomment=[s]{/*}{*/},
  morestring=[b]",                      
  morekeywords={
    ref,%
    void,%
    for,%
    if,%
    then,%
    else,%
    end,%
    return,%
    function,%
    while,%
    repeat,%
    until,%
    foreach,%
    do,%
    min,%
    max,%
    switch,%
    case},
  identifierstyle=,                      
  keywordstyle=[1]\color{black!50!brown}\bf, 
  commentstyle=\usefont{T1}{pcr}{m}{sl}\color{green!50!black}\small, 
  stringstyle=\color{purple}\small,       
  emphstyle={\color{red}\bfseries},       
  emphstyle={[2]\color{purple}\bfseries}, 
  emphstyle={[3]\color{blue}},            
  showstringspaces=false,                 
  numbers=left,                           
  firstnumber=1,                          
  numberstyle=\tiny\color{blue},          
  breaklines=true,
  breakatwhitespace=true,
  emptylines=1,
  texcl=true,                             
  mathescape=true,                        
  escapeinside={@}{@},
  xleftmargin=2.5ex,
  autogobble
}

\let\oldFootnote\footnote
\newcommand\nextToken\relax
\renewcommand\footnote[1]{%
    \oldFootnote{#1}\futurelet\nextToken\isFootnote}
\newcommand\isFootnote{%
    \ifx\footnote\nextToken\textsuperscript{,}\fi}

\makeatletter
\DeclareFontFamily{OMX}{MnSymbolE}{}
\DeclareSymbolFont{MnLargeSymbols}{OMX}{MnSymbolE}{m}{n}
\SetSymbolFont{MnLargeSymbols}{bold}{OMX}{MnSymbolE}{b}{n}
\DeclareFontShape{OMX}{MnSymbolE}{m}{n}{
  <-6>  MnSymbolE5
  <6-7>  MnSymbolE6
  <7-8>  MnSymbolE7
  <8-9>  MnSymbolE8
  <9-10> MnSymbolE9
  <10-12> MnSymbolE10
  <12->   MnSymbolE12
}{}
\DeclareFontShape{OMX}{MnSymbolE}{b}{n}{
  <-6>  MnSymbolE-Bold5
  <6-7>  MnSymbolE-Bold6
  <7-8>  MnSymbolE-Bold7
  <8-9>  MnSymbolE-Bold8
  <9-10> MnSymbolE-Bold9
  <10-12> MnSymbolE-Bold10
  <12->   MnSymbolE-Bold12
}{}

\let\llangle\@undefined
\let\rrangle\@undefined
\DeclareMathDelimiter{\llangle}{\mathopen}%
{MnLargeSymbols}{'164}{MnLargeSymbols}{'164}
\DeclareMathDelimiter{\rrangle}{\mathclose}%
{MnLargeSymbols}{'171}{MnLargeSymbols}{'171}
\makeatother

\ifpdf
  \DeclareGraphicsExtensions{.eps,.pdf,.png,.jpg}
\else
  \DeclareGraphicsExtensions{.eps}
\fi

\usepackage[capitalize]{cleveref}

\crefname{subsection}{Section}{Section}
\crefname{subsubsection}{Section}{Section}
\Crefname{subsection}{Section}{Section}
\Crefname{subsubsection}{Section}{Section}

\usepackage{showkeys}

\newcommand{\Vb}{{\mathbf{b}}}

\newcommand{\Vf}{{\mathbf{f}}}

\newcommand{\Vj}{{\mathbf{j}}}

\newcommand{\Vp}{{\mathbf{p}}}
\newcommand{\Vq}{{\mathbf{q}}}

\newcommand{\Vu}{{\mathbf{u}}}
\newcommand{\Vv}{{\mathbf{v}}}
\newcommand{\Vw}{{\mathbf{w}}}

\newcommand{\Vzero}{\mathbf{0}}

\newcommand{\VA}{{\mathbf{A}}}
\newcommand{\VB}{{\mathbf{B}}}

\newcommand{\VD}{{\mathbf{D}}}
\newcommand{\VE}{{\mathbf{E}}}

\newcommand{\VH}{{\mathbf{H}}}

\newcommand{\VP}{{\mathbf{P}}}

\newcommand{\VW}{{\mathbf{W}}}

\newcommand{\BC}{{\boldsymbol{C}}}

\newcommand{\BH}{{\boldsymbol{H}}}

\newcommand{\BL}{{\boldsymbol{L}}}

\newcommand{\Bn}{{\boldsymbol{n}}}

\newcommand{\Bs}{{\boldsymbol{s}}}
\newcommand{\Bt}{{\boldsymbol{t}}}

\newcommand{\Bx}{{\boldsymbol{x}}}

\newcommand{\etabf}{\boldsymbol{\eta}}

\newcommand{\iotabf}{\boldsymbol{\iota}}

\newcommand{\nubf}{\boldsymbol{\nu}}

\newcommand{\pibf}{\boldsymbol{\pi}}

\newcommand{\rhobf}{\boldsymbol{\rho}}

\newcommand{\sigmabf}{\boldsymbol{\sigma}}

\newcommand{\phibf}{\boldsymbol{\phi}}
\newcommand{\varphibf}{\boldsymbol{\varphi}}

\newcommand{\omegabf}{\boldsymbol{\omega}}

\newcommand{\Phibf}{\boldsymbol{\Phi}}

\newcommand{\Ca}{{\cal A}}

\newcommand{\Ce}{{\cal E}}

\newcommand{\Cl}{{\cal L}}

\newcommand{\Cv}{{\cal V}}
\newcommand{\Cw}{{\cal W}}

\newcommand{\bbR}{\mathbb{R}}

\newcommand*{\Op}[1]{\mathsf{#1}} 
\newcommand*{\blf}[1]{\mathsf{#1}} 

\providecommand*{\wt}[1]{\widetilde{#1}}
\providecommand*{\wh}[1]{\widehat{#1}}

\providecommand*{\N}[1]{\left\|{#1}\right\|} 

\providecommand{\Div}{\operatorname{div}}          
\providecommand{\curl}{\operatorname{{\bf curl}}}  
\providecommand{\grad}{\operatorname{{\bf grad}}}       

\providecommand{\Supp}{\operatorname{supp}}                            
\providecommand{\supp}{\Supp}

\providecommand{\Id}{\Op{Id}}                     

\newcommand{\defaultdomain}{\Omega}
\newcommand{\defaultboundary}{\partial\defaultdomain}

\providecommand*{\Lp}[2][\defaultdomain]{L^{#2}({#1})}
\newcommand*{\Lpv}[2][\defaultdomain]{\BL^{#2}({#1})}

\newcommand*{\Ltwo}[1][\defaultdomain]{\Lp[#1]{2}}
\newcommand*{\Ltwov}[1][\defaultdomain]{\Lpv[#1]{2}}

\NewDocumentCommand{\Hm}{O{\defaultdomain}md<>}{%
\IfValueTF{#3}{H^{#2}_{#3}(#1)}{H^{#2}(#1)}}            

\newcommand*{\bHm}[3][\defaultdomain]{H_{#3}^{#2}({#1})}

\newcommand*{\zbHone}[1][\defaultdomain]{\bHm[#1]{1}{0}}

\newcommand{\hlb}{\frac{1}{2}}
\NewDocumentCommand{\Hh}{O{\defaultboundary}d<>}{%
\IfValueTF{#2}{\bHm[#1]{\hlb}{#2}}{\Hm[#1]{\hlb}}}

\NewDocumentCommand{\Hmh}{O{\defaultboundary}d<>}{%
\IfValueTF{#2}{\bHm[#1]{-\hlb}{#2}}{\Hm[#1]{-\hlb}}}

\newcommand*{\zbHcurl}[1][\defaultdomain]{\bHcurl[#1]{0}}

\newcommand{\contspcsymb}{C}
\NewDocumentCommand{\Contm}{sd<>O{\defaultdomain}m}{%
  \IfBooleanTF{#1}{\renewcommand{\contspcsymb}{\BC}}{\renewcommand{\contspcsymb}{C}}%
  \IfValueTF{#2}{\contspcsymb_{#2}^{#4}({#3})}{\contspcsymb^{#4}({#3})}}

\DeclareDocumentCommand{\itg}{s O{\defaultdomain} o m O{\mathrm{d}\Bx}}{%
  \IfBooleanTF{#1}%
  {\IfNoValueTF{#3}{\int\nolimits_{#2}{#4}\,{#5}}{\int\nolimits_{#2}^{#3}{#4}\,{#5}}}%
  {\IfNoValueTF{#3}{\int\limits_{#2}{#4}\,{#5}}{\int\limits_{#2}^{#3}{#4}\,{#5}}}}%

\definecolor{darkgreen}{rgb}{0,0.7,0}
\definecolor{darkyellow}{rgb}{1.0,0.75,0.0}

\newcommand*{\cob}[1]{{\color{blue}{#1}}}

\newcommand*{\cop}[1]{{\color{purple}{#1}}}

\NewDocumentCommand{\samvismat}{r()r()om}{%
    \IfValueTF{#3}{%
      \left[%
      \begin{array}{c}%
      \psframebox[fillstyle=solid,fillcolor=#3,linecolor=white,linewidth=0pt,framesep=0pt]%
      {\parbox[c][#2][c]{#1}{\centering{#4}}}%
      \end{array}%
      \right]%
    }{%
  \left[%
      \begin{array}{c}%
      {\parbox[c][#2][c]{#1}{\centering{#4}}}%
      \end{array}%
      \right]%
    }%
}

\NewDocumentCommand{\samvisvec}{r()om}{%
        \typeout{samvisvec #1}
    \IfValueTF{#2}{%
      \left[%
      \begin{array}{c}%
      \psframebox[fillstyle=solid,fillcolor=#2,linecolor=#2,linewidth=0pt,framesep=0pt]%
         {\begin{array}[c]{c}\rule{0pt}{#1}\end{array}%
           \begin{array}[c]{c}{#3}\end{array}}     
      \end{array}%
      \right]%
    }{%
  \left[%
      \begin{array}{c}%
             {\begin{array}[c]{c}\rule{0pt}{#1}\end{array}}     
             {\begin{array}[c]{c}{#3}\end{array}}     
      \end{array}%
      \right]%
    }%
}

\NewDocumentCommand{\samterm}{s O{red} m}{%
  \IfBooleanTF{#1}{{\color{#2}\textbf{#3}}}%
  {{\color{#2}{#3}}}%
}

\NewDocumentCommand{\samemp}{s O{magenta} m}{%
  \IfBooleanTF{#1}{{\color{#2}\textit{#3}}}%
  {{\color{#2}{#3}}}%
}

\NewDocumentCommand{\samhl}{O{yellow!50!white} m}{%
  \psframebox[framesep=0.5ex,fillstyle=solid,fillcolor=#1,linecolor=#1]{#2}
}

\newcommand{\dom}{\Omega}

\newcommand{\extd}{\operatorname{\mathsf{d}}}
\newcommand{\ecd}{\extd^{*}}
\newcommand{\fen}{\Ce_{\textrm{loc}}}
\newcommand{\wen}{\Cw}
\newcommand{\lml}{\Op{M}_{\textrm{loc}}}
\newcommand{\gml}{\Op{M}}

\newcommand{\Hdf}[1][\dom]{H\Lambda^{\ell}(\extd,#1)}
\newcommand{\velo}{\boldsymbol{\Cv}}
\newcommand{\rfx}{\wh{\Bx}}
\newcommand{\Derv}{\Op{D}}
\newcommand{\hop}{\star}
\newcommand{\Lie}[1][\velo]{\Op{L}_{#1}}
\newcommand{\contr}[1][\velo]{\operatorname{\iotabf}_{\velo}}
\newcommand{\froz}[1]{\overline{#1}}
\newcommand{\Aloc}{\Ca_{\mathrm{loc}}}
\newcommand{\wenl}{\wen_{L}}
\newcommand{\mycolorbox}[3][yellow!30!white]{\fcolorbox{#1}{#1}{\parbox{#2}{#3}}}
\newcommand{\vpr}[1]{\overrightarrow{#1}}

\theoremstyle{definition}
\newtheorem{remark}[theorem]{Remark}

\newcommand{\skcomment}[1]{}

\headers{EM Force by Shape Calculus}{P. Panchal, R. Hiptmair, and S. Kurz}
\title{``Holding the Fields Constant''\\
A Shape-Calculus Approach to Electromagnetic Forces} 
\author{P. Panchal\thanks{SAM, ETH Zurich, CH-8092 Z\"urich,}
  \email{piyush.panchal\symbol{64}sam.math.ethz.ch} \and 
  R. Hiptmair\thanks{SAM, ETH Zurich, CH-8092 Z\"urich,}
  \email{ralf.hiptmair\symbol{64}sam.math.ethz.ch} \and 
  S.~Kurz\thanks{SAM, ETH Zurich, CH-8092 Z\"urich,}
  \email{stefan.kurz\symbol{64}sam.math.ethz.ch}
}
\date{\today}

\begin{document}

\maketitle



\section*{Abstract}
\subsection*{Purpose}
The purpose of this paper is to perform a rigorous investigation of the virtual work
principle for the computation of electromagnetic forces in static settings using the
mathematical theory of shape calculus.
\subsection*{Design/methodology/approach}
Our approach is motivated by the widely held belief that the virtual work principle (VWP)
for modeling electromagnetic forces should be applied with passively advecting the
fields with the virtual displacement.  This study employs the mathematical theory of shape
calculus as a tool in the VWP framework. The adjoint approach to the differentiation of
functionals under variational constraints is utilized to provide a mathematical
justification for the computation of deformation-dependent field energy gradients.
\subsection*{Findings}
The study reveals that passively advecting the actual electromagnetic fields is appropriate
only in the case of linear materials. For general cases, it is necessary to consider
fields arising as solutions of adjoint variational problems.
\subsection*{Research limitations/implications}
This paper focuses on static settings and the investigation is limited to the theoretical
framework of shape calculus. Future work could extend the analysis to dynamic settings and
explore the application to actual force computation with the finite element method in the
case of non-linear materials.
\subsection*{Originality/value}
The rigorous application of shape calculus to the virtual work principle in
electromagnetic force computation is novel. This study provides a mathematical
justification for the widely held belief about field advection when applying the VWP,
contributing to a deeper understanding of electromagnetic forces, particularly in
non-linear material contexts, where the solution of an adjoint variational problem has to
be considered, too.
\medskip

\textbf{Keywords.} Shape calculus, electromagnetic forces, virtual work principle, adjoint
variational problems

\section{Introduction}
\label{sec:intro}

\subsection{Virtual work principle}
\label{ss:vwp}

Electromagnetic forces effect the transfer of energy between the electromagnetic and
mechanical sub-systems of a physical system, see, for example, \cite{HEH07} and
\cite[Section~I]{BOS05}. This is reflected in the fundamental \emph{virtual work
  principle}, which views the electromagnetic force field $\Vf$ in the direction of a
\emph{deformation vector field} $\velo$ as the derivative of the field energy. Writing
$\Omega\subset\bbR^{3}$ for the field domain, we can cast this consideration into the
pseudo-formula
\begin{gather}
  \label{eq:40}
  \text{``}\quad
  \int\nolimits_{\Omega}\Vf(\Bx)\cdot\velo(\Bx)\,\mathrm{d}\Bx = \frac{d\,\{\text{field
      energy\}}}{d\,\{\text{deformation in the direction of $\velo$\}}}\quad \text{''}
  \;.
\end{gather}
\emph{Shape calculus} is the field of applied mathematics that gives a rigorous meaning to such a
derivative. Below, from \Cref{sec:vwm} we will introduce and use it.

\subsection{``Holding the fields constant''}
\label{ss:hfc}

The field energy as in \eqref{eq:40} depends on the geometric configuration. Hence, it
changes as that geometric configuration is deformed along $\velo$.  This change can be
partly attributed to the deformation of the medium/domain, but another contribution to the
change in field energy is that the electromagnetic fields depend on the geometric
configuration.

The starting point for this article was the work \cite{HEH07a} by F.~Henrotte and
K.~Hameyer, where in Section~III in the context of a magnetostatic model problem the
authors propose the formula, here rendered with adjusted notation,
\begin{gather}
  \label{eq:41}
  \frac{d\,\{\text{field
      energy}\}}{d\,\{\text{deformation in the direction of $\velo$}\}} =
  - \left.\frac{\partial \wen_{M}}{\partial s}\right|_{ \frac{D}{D\velo}\VB=0} \;.
\end{gather}
In this formula $s$ is the ``deformation parameter''\footnote{The deformation control
  parameter $s$ must not be mistaken for a ``time'', hence the notation.} for the flow
induced by $\velo$, and $\wen_{M}$ is the magnetic field energy. The important ingredient
in \cref{eq:41} is the constraint $ \frac{D}{D\velo}\VB=0$, implying that the material
derivative of the magnetic flux density $\VB$ (viewed as a 2-form) with respect to $\velo$
has to vanish. In physical terms this means that the $s$-dependence of $\VB$ must be such
that the magnetic flux through any surface remains constant as that surface moves with
$\velo$: $\frac{\partial\VB}{\partial s} + \curl(\velo\times\VB)=0$\footnote{In the
  language of differential geometry this expresses
  $\frac{\partial\VB}{\partial t}+\Op{L}_{\velo}\VB = 0$, where $\Op{L}_{\velo}$ denotes
  the Lie derivative for 2-forms.}. This is the recipe for taking into account the
deformation-dependence of the $\VB$-field. No further justification is given in
\cite{HEH07a}.

This recipe is invoked in other works and is usually taken for granted. In fact, it is
universally employed in the work of J.L.~Coulomb \cite{CMS82,COU83}, which the authors
acknowledge to be based on the 1977 thesis of P.~Rafinejad \cite{RAF77}, where we can read,
without further explanation or a reference,
\begin{quote}
  En effet, il s'agit d'une variation en fonction du d\'eplacement en gardant l'autre
  variable ind\'ependante (le potentiel) constante.
\end{quote}
In these works, the ``holding the fields constant'' technique is applied in combination
with finite-element Galerkin discretization. In this context it means that in the course
of a virtual deformation the finite-element degrees of freedom retain their values and the
computation of the derivative \cref{eq:41} is called the ``local Jacobian derivative
method''.

Then we realized that throughout the work of A.~Bossavit, an eminent researcher as regards
modeling of electromagnetic forces, the ``holding the fields constant'' paradigm plays a
central role, as in \cite[Section~4.2]{BOS92}, where he writes that
\begin{quote}
  force $\ldots$ is simply the \emph{partial derivative of} $\Phi$ (the co-energy)
  \emph{with respect to} $u$ (the configuration): $\Vf = \frac{\partial \Phi}{\partial u}(u,\VH)$.
\end{quote}
That partial derivative encodes ``holding $\VH$ constant'' when computing the derivative,
similarly as in \eqref{eq:41}.  In \cite[Section~I]{BOS90f}, \cite[Section~IV]{BOS92f},
and \cite[Section~III]{BOS05} we find a variation of this theme:
\begin{quote}
  ``force is obtained by differentiating the coenergy with respect to the configuration
  parameter, \emph{while keeping currents constant}.''
\end{quote}
That last paper \cite{BOS05} and \cite{BOS11} also give a ``physics-inspired proof'' of
this claim appealing to energy conservation and the usual formula for Joule losses. This
comes closest to the sought-after mathematical foundation for the ``keeping the fields
constant'' approach in the framework of the virtual work principle.

``Holding the fields constants'' is employed also beyond computational engineering. For
instance, in \cite{SVA06} the authors repeatedly mention doing ``virtual displacement''
under constant currents, field or flux densities, depending on the model: A magnetic
scalar potential is treated this way in Formula (23), constant magnetic flux is discussed
in Section~3.2.

\subsection{Outline and main results}
\label{ss:out}

The purpose of this article is to make precise and rigorously justify \cref{eq:41}.  The
focus is on electrostatic and magnetostatic fields and closed systems, though our
considerations can probably be extended to quasi-static settings. The corresponding
boundary value
problems are introduced in \Cref{ss:stat}, first in classical
vector-analytic notation, then in the language of exterior calculus (differential
forms). We eschew full generality by confining ourselves to local material laws. Based on
field energies we state variational formulations in \Cref{ss:vf}. \Cref{sec:vwm} reviews
the virtual work principle from the perspective of shape calculus. This is complemented by
Sections~\ref{ss:adjm} and \ref{ss:asma}, which discuss the so-called adjoint approach to
differentiation under variational constraints.
 
After these preparations, in \Cref{sec:force} we harness the adjoint approach to
shape-differentiate field energies, which gives us their shape derivatives with respect to
a deformation vector field $\velo$, the right-hand side of \cref{eq:40}. The main result
is \cref{eq:main}, which gives precise mathematical meaning to the passive advection
recipe in the application of the virtual work principle. However, that formula also
contains an auxiliary field obtained as the solution of an adjoint variational problem.
In the end of \Cref{ss:sdfe} we manipulate \cref{eq:main} further to obtain more explicit
formulas for the force density. 

In \Cref{ss:vpff} we recast these expressions using classical vector calculus. Finally, in
\Cref{sec:lin} we prove that for \emph{linear materials} one can dispense with solving adjoint
variational problems. Thus we arrive at the ``recipe'' promoted in much literature. It is
rigorously captured in \cref{eq:28}.

\section{Setting and Model Problems}
\label{sec:set}

Throughout we write $\dom\subset\bbR^{3}$ for a domain with Lipschitz boundary, which may
be bounded or not; even $\dom=\bbR^{3}$ is admitted. 

\subsection{Static electromagnetic fields}
\label{ss:stat}

In a static setting Maxwell's equations can be split into two independent sets of
equations. Using the standard notations for the various electromagnetic fields, those read
\begin{align}
  \label{eq:es}
  & \text{for electrostatics:}  & \curl\VE & = 0\;, &  \Div\VD & = \rho_{i} \;,\\
  \label{eq:ms}
  & \text{for magnetostatics:} & \Div\VB & = 0 \;, &  \curl\VH & = \Vj_{i} \;.
\end{align}
The functions $\rho_{i}$ and $\Vj_{i}$ represent given sources, charge densities and
impressed currents, respectively. They must have compact support and, in addition, we
demand, $\Div\Vj_{i}=0$. The left equations of \cref{eq:es} and \cref{eq:ms} can be
expressed by means of potentials, yielding the equivalent PDEs,
\begin{align}
  \label{eq:ep}
  & \text{for electrostatics:}  & \VE & = -\grad V\;, &  \Div\VD & = \rho_{i} \;,\\
  \label{eq:mp}
  & \text{for magnetostatics:} & \VB & = \curl\VA \;, &  \curl\VH & = \Vj_{i} \;,
\end{align}
where $V$ is the electric scalar potential and $\VA$ the magnetic vector potential. We
impose zero traces on $\partial\Omega$ and decay conditions at infinity:
\begin{align}
  \label{eq:ec}
  & \text{electrostatics:}  &  V & = 0 \quad\text{on}\;\partial\dom\;, &  V(\Bx) & =
  O(\N{\Bx}^{-1})\quad\text{for}\;\N{\Bx}\to\infty \;,\\
  \label{eq:mc}
  & \text{magnetostatics:} & \VA_{\Bt} & = \Vzero \quad\text{on}\;\partial\dom\;, & \VA(\Bx) & =
  O(\N{\Bx}^{-1})\quad\text{for}\;\N{\Bx}\to\infty \;.
\end{align}
Here $\VA_{\Bt}$ stands for the tangential component of $\VA$ on $\partial \dom$. 

The equations \cref{eq:es} and \cref{eq:ms} have to be supplemented with material
laws\footnote{The signs reflect widely used conventions in the theory of electromagnetic fields.}
\begin{gather}
  \label{eq:ml}
  \VD = -\Op{M}_{\mathrm{e}}(-\VE) \quad , \quad \VH = \Op{M}_{\mathrm{m}}(\VB) \;.
\end{gather}
The mappings\footnote{We use the customary notation $\Ltwo$ for square-integrable
  functions on $\dom$; bold symbols indicate spaces of vector fields.}
$\Op{M}_{*}:\Ltwov \to \Ltwov$, $*=\mathtt{e},\mathrm{m}$, are supposed to be strongly
monotone, in particular, bijective \cite[Section~25.3]{ZEI90m}. Material laws are
\emph{local}, when \cref{eq:ml} can be written as
\begin{gather}
  \label{eq:mlloc}
  \VD(\Bx) = -\Op{M}_{\mathrm{e},\mathrm{loc}}(-\VE(\Bx)) \quad , \quad
  \VH(\Bx) = \Op{M}_{\mathrm{m},\mathrm{loc}}(\VB(\Bx)) \;,\quad \Bx\in \dom \;,
\end{gather}
with strongly monotone mappings $\Op{M}_{*,\mathrm{loc}}:\bbR^{3}\to \bbR^{3}$. The
simplest case are local, \emph{linear}, and isotropic material laws
\begin{gather}
  \label{eq:mlll}
  \VD(\Bx) = \epsilon(\Bx)\VE(\Bx) \quad,\quad
  \VH(\Bx) = \chi(\Bx)\VB(\Bx)\;,\quad \Bx\in\dom \;,
\end{gather}
with uniformly positive and bounded functions $\epsilon,\chi:\dom\to\bbR^{+}$.

Combining \cref{eq:ep}, \cref{eq:mp}, the corresponding material laws \cref{eq:ml}, and
taking into account \cref{eq:ec}, \cref{eq:mc}, respectively, we arrive at the boundary
value problems
\begin{align}
  \label{eq:ebvp}
   \text{electrostatics:}\quad  &
  \left\{\begin{gathered}
      -\Div \Op{M}_{\mathrm{e}}(\grad V)  = \rho_{i} \quad\text{in}\quad \dom \;,\\
      V = 0 \quad \text{on}\;\partial\dom\;,\quad  V(\Bx) =
      O(\N{\Bx}^{-1})\quad\text{for}\;\N{\Bx}\to\infty \;,
  \end{gathered}\right.
\\
  \label{eq:mbvp}
  \text{magnetostatics:}\quad &
  \left\{
    \begin{gathered}
      \curl\Op{M}_{\mathrm{m}}(\curl\VA)  = \Vj_{i} \quad \text{in}\quad \dom \;,\\
     \VA_{\Bt}  = \Vzero \quad\text{on}\;\partial\dom\;,  \VA(\Bx)  =
    O(\N{\Bx}^{-1})\quad\text{for}\;\N{\Bx}\to\infty \;.
  \end{gathered}
\right.
\end{align}

\subsection{Exterior calculus perspective}
\label{ss:ecp}
We adhere to A.~Bossavit's tenet \cite{BOS92,BOS08a,BOS91b,BOS98,BOS98a,BOS98b,BOS98c,BOS98d} that 
Exterior calculus offers a unifying way to state the boundary value problems
\cref{eq:ebvp} and \cref{eq:mbvp}, refer to \cite{WAR14}, \cite[Section~2]{HIP02} or
\cite[Section~2.2]{HIP15}. We seek a differential form\footnote{We use bold Greek symbols
  for differential forms.}  $\varphibf$ of degree $\ell$ on $ \Omega$, that is, a mapping
from $\Omega$ into the space $\Lambda^{\ell}(\bbR^{3})$ of alternating $\ell$-multilinear
forms on $\bbR^{3}$, satisfying
\begin{gather}
  \label{eq:fbvp}
  \boxed{\begin{gathered}
    \ecd\Op{M}(\extd \varphibf) =  \sigmabf \quad\text{in}\quad \Omega \;,\\
    \varphibf_{\Bt} = 0 \quad\text{on }\partial\Omega\;,\quad \N{\varphi(\Bx)} =
    O(\N{\Bx}^{-1}) \quad\text{for}\;\N{\Bx}\to\infty \;,
  \end{gathered}}
\end{gather}
where $\extd$ denotes the \emph{exterior derivative} acting on $\ell$-forms, $\ecd$
(acting on $2-\ell$-forms), stands for its adjoint with respect to the
$L^{2}(\Omega)$-type $\wedge$-pairing of $k$-forms and $3-k$-forms, and $\varphibf_{\Bt}$
designates the \emph{tangential trace} of $\varphi$ on $\partial\Omega$. Further, in
\cref{eq:fbvp} the material law $\Op{M}$ has to be read as an operator
$L^{2}\Lambda^{\ell+1}(\Omega)\to L^{2}\Lambda^{2-\ell}(\Omega)$, where we write
$L^{2}\Lambda^{k}(\Omega)$ for the Hilbert space of square-integrable $k$-forms on
$\Omega$ \mbox{$ $}\footnote{This notation for (Sobolev) spaces of differential forms is borrowed
  from \cite[Section~2.1]{AFW06}.}. Finally, $\sigmabf\in L^{2}\Lambda^{3-\ell}(\Omega)$
is a given compactly supported ``source'' $3-\ell$-form on $\Omega$, with the
``orthogonality property''
\begin{gather}
  \label{eq:ko}
  \int\nolimits_{\Omega}\sigmabf\wedge\etabf = 0 \quad \forall \etabf\in
  L^{2}\Lambda^{\ell}(\Omega),\; \etabf_{\Bt}=0\;\text{on }\partial\Omega,\;  \extd\etabf=0 \;,
\end{gather}
which is necessary to ensure solvability of \cref{eq:fbvp}. 

Employing the canonical identification of differential forms with their \emph{Euclidean
  vector proxies} in $\bbR^{3}$ \cite[Section~2.2.2]{HIP15}, from \cref{eq:fbvp} we
recover \cref{eq:ebvp} (with $V \leftrightarrow \varphibf$) for $\ell=0$ and
\cref{eq:mbvp} (with $\VA \leftrightarrow \varphibf$) for $\ell=1$ .

\subsection{Field energies}
\label{ss:fe}

In the framework of \cref{eq:fbvp} we rely on the following purely local
expression for the \emph{field energy} \cite[Section~2.4.2]{HIP15}
\begin{gather}
  \label{eq:wen}
  \wen(\varphibf,
  \rhobf) :=
  \int\nolimits_{\Omega}\fen(\Bx,(\extd \varphibf)(\Bx),\rhobf(\Bx))
  \;,\quad
  \extd \varphibf\in L^{2}\Lambda^{\ell+1}(\Omega)\;,\quad \rhobf \in L^{1}\Lambda^{3}(\Omega) \;,
\end{gather}
$\Bx$ a point in $\Omega$, based on a local field energy density
\begin{gather}
  \label{eq:fen}
  \fen: \left\{
    \begin{array}[c]{ccl}
      \Omega \times \Lambda^{\ell+1}(\bbR^{3}) \times \Lambda^{3}(\bbR^{3}) &\to &
      \Lambda^{3}(\bbR^{3})\;, \\
      (\Bx,\omega,\rho) &\mapsto & \fen(\Bx,\omega,\rho)\;.
    \end{array}
  \right. 
\end{gather}
We expect $\fen(\Bx,\omega,\rho)$ to be a non-negative multiple of the volume form in
$\Bx\in\Omega$.  In \cref{eq:wen} $\varphibf$ is the solution of \cref{eq:fbvp}, whereas
$\rhobf$ is a given $3$-form, describing the spatially varying density of the
medium\footnote{We also stipulate that $\rhobf(\Bx)$ is a non-negative multiple of the
  volume form for almost all $\Bx\in\Omega$.} \cite{BOS17}. We demand that
$\fen = \fen(\Bx,\omega,\rho)$ is continuously differentiable in the second argument
$\omega$ and that there are constants $0<c^{-}\leq c^{+}<\infty$ such
that\footnote{Throughout this article we use angle brackets $\left<\cdot,\cdot\right>$ to
  denote duality pairings, that is, the evaluation of a linear form (on the left) for some
  argument (on the right).}\textsuperscript{,}\footnote{Recall that the derivative in
  $x\in X$ of a differentiable mapping $X\mapsto Y$, $X$, $Y$ (real) Banach spaces, is a
  bounded linear operator $X\to Y$. Hence, the derivative of an $\bbR$-valued mapping
  $X\to \bbR$ in $x\in X$ is a bounded linear form on $X$.}
\begin{align}
  \label{eq:mon}
  \hop \left<
    \frac{\partial \fen}{\partial \omega}(\Bx,\omega,\rho) - \frac{\partial \fen}{\partial
      \omega}(\Bx,\omega',\rho),
    \omega-\omega'
  \right> & \geq c^{-}\N{\omega-\omega'}^{2} \;,\\
  \label{eq:lcont}
  \left|\hop
    \left<\frac{\partial \fen}{\partial \omega}(\Bx,\omega,\rho) - \frac{\partial \fen}{\partial
      \omega}(\Bx,\omega',\rho),
    \omega-\omega'
  \right>\right| & \leq c^{+}\N{\omega-\omega'}^{2} 
\end{align}
for all $\omega,\omega'\in \Lambda^{\ell+1}(\bbR^{3})$ and for all
$\Bx\in\Omega$, $\rho\in \Lambda^{3}(\bbR^{3})$. Here $\hop$ is the Euclidean Hodge
operator $\hop$ \cite[XV,\S 4]{LAN99}. This implies that $\fen$ is strongly
convex in the second argument and that $|\fen(\Bx,\omega,\rho)| = O(\N{\omega}^{2})$
uniformly for $\N{\omega}\to\infty$.

The energy density induces a \emph{local material law} $\lml:\Omega\times
\Lambda^{\ell+1}(\bbR^{3})\times \Lambda^{3}(\bbR^{3})\to \Lambda^{2-\ell}(\bbR^{3})$
through
\skcomment{
1.  Is it that $\partial \Ce / \partial \omega$ behaves like a (l+1)-vector? This should be the case, in the light of the pairing. How can this be formally justified?

2. We should avoid the Euclidean Hodge here. The metrical part of the model should be entirely encapsulated in the material operator. There is no justification for an additional Euclidean background metric.

3. The first step to avoid the Hodge is introducing a volume form, say "dx". Then, we can skip the Hodge and add the volume form on the RHS. But where to get the volume form from?

4. The answer lies in (2.14), where "dx" is required to integrate the energy density. "dx" is unique up to a positive function, and this function cancels out in (2.18).

5. This function could be avoided by defining the energy density as 3-form. Then
$\partial \Ce / \partial \omega$ could be modeled as 3-form that takes values in the
space of (l+1)-vectors (physics language: mixed tensor). Then, the Hodge in (2.28) is
gone. In fact, $M_{loc}$ is just the self-contraction of $\partial \Ce / \partial \omega$.

6. In the differential form calculus, this would come out automatically like this. In the
linear case, $\Ce = 1/2 E^D + 1/2 B^H$.
}
\begin{gather}
  \label{eq:lml}
  \lml(\Bx,\omega,\rho)\wedge \nu = \left<\frac{\partial \fen}{\partial
      \omega}(\Bx,\omega,\rho),\nu\right>\quad\forall \nu\in \Lambda^{\ell+1}(\bbR^{3})
  \;.
\end{gather}
In turn, \cref{eq:lml} gives rise to the global material law
\begin{gather}
  \label{eq:glm}
  \left\{
  \begin{aligned}
    & \gml:L^{2}\Lambda^{\ell+1}(\Omega)\times L^{1}\Lambda^{3}(\Omega) \to
    L^{2}\Lambda^{2-\ell}(\Omega)\;,\\
    & \gml(\omegabf,\rhobf)(\Bx) := \lml(\Bx,\omegabf(\Bx),\rhobf(\Bx))\;,\quad\Bx\in \Omega\;.
  \end{aligned}
  \right.
\end{gather}
The monotonicity property \cref{eq:mon} implies that
${\frac{\partial\wen}{\partial\omegabf}}$ is a monotone operator. Hence, for all
$\rhobf\in L^{1}\Lambda^{3}(\bbR^{3})$, $\omegabf\mapsto \gml(\omegabf,\rhobf)$ is a
continuous and bijective mapping
$L^{2}\Lambda^{\ell+1}(\Omega)\to L^{2}\Lambda^{2-\ell}(\Omega)$ with continuous inverse
$\gml^{-1}(\cdot,\rhobf)$ \cite[Section~25.4]{ZEI90m}.

\subsection{Variational formulations}
\label{ss:vf}

The (primal) weak form of \cref{eq:fbvp} seeks\footnote{
  In case $\Omega$ is unbounded, we have to resort to the weighted spaces
  \begin{gather*}
    L^{2}_{w}\Lambda^{k}(\Omega) := \left\{
      \Bx \mapsto \frac{1}{\sqrt{1+\N{\Bx}^{2}}} \nubf(\Bx) \in L^{2}\Lambda^{k}(\Omega)
    \right\}\;,\quad k \in\{0,\ldots,3\}\;,
  \end{gather*}
  to capture the decay condition involved in \cref{eq:fbvp}.
}\skcomment{I do "believe" this - very plausible, I have not checked the details.}
\begin{gather}
  \label{eq:Hdf}
  \varphibf \in \Hdf := \{\nubf\in L_{w}^{2}\Lambda^{\ell}(\Omega):\, \extd\nubf\in
  L^{2}\Lambda^{\ell+1}(\Omega),\, \nubf_{\Bt}=0\;\text{on }\partial\Omega\}  
\end{gather}
such that \cite{HEH07}
\begin{gather}
  \label{eq:vff}
  \left<
    \frac{\partial\wen}{\partial\varphibf}(\varphibf,\rhobf),\etabf
  \right> = 
  \int\nolimits_{\Omega} \gml(\extd \varphibf,\rhobf) \wedge \extd \etabf =
  \int\nolimits_{\Omega}\sigmabf\wedge \etabf\quad \forall \etabf\in \Hdf \;.
\end{gather}
This equation characterizes a stationary point of the field Lagrangian
\begin{gather}
  \label{eq;fL}
  \etabf \in \Hdf \mapsto
  \wen(\etabf,\rhobf)  - \int\nolimits_{\Omega}\sigmabf\wedge \etabf \;.
\end{gather}
Both, the density 3-form $\rhobf$ and the source field
$\sigmabf\in L^{2}\Lambda^{3-\ell}(\Omega)$ can be regarded as given.  Using
\eqref{eq:wen} and \cref{eq:lml}, we can also rewrite \cref{eq:vff} in the more detailed
local form: Seek $\varphibf\in\Hdf$ such that
\begin{gather}
  \label{eq:vffl}
  \boxed{\int\nolimits_{\Omega}
  \left<
    \frac{\partial \fen}{\partial\omega}(\Bx,(\extd \varphibf)(\Bx),\rhobf(\Bx)),\extd
    \etabf(\Bx)
  \right>
  =  \int\nolimits_{\Omega}\sigmabf\wedge \etabf \quad\forall \etabf\in\Hdf} \;,
\end{gather}
or, using \cref{eq:lml},
\begin{gather}
  \label{eq:vfflm}
  \int\nolimits_{\Omega}
  \lml(\Bx,(\extd \varphibf)(\Bx),\rhobf(\Bx)) \wedge \extd
    \etabf(\Bx)
  =  \int\nolimits_{\Omega}\sigmabf\wedge \etabf \quad\forall \etabf\in\Hdf \;.
\end{gather}
 Thanks to \cref{eq:ko}, \cref{eq:mon} and \cref{eq:lcont} existence of solutions of
 \cref{eq:vff}/\cref{eq:vffl} follows from the theory of monotone operators
 \cite[Theorem~25.B]{ZEI90m}. We point out that \cref{eq:vffl} is an \emph{ungauged}
 variational formulation, thanks to \cref{eq:ko} with a consistent right-hand side.
 Therefore, the solution for the potential $\varphibf$ may not be unique. However, the
 field $\extd \varphibf$ will be unique. Since it is only that field that enters the
 expression \cref{eq:wen} for the field energy, this one will also be uniquely determined.
 
\section{The Virtual Work Method and Shape Calculus}
\label{sec:vwm}

We elaborate shape differentiation by the velocity method of shape calculus, see
\cite[Section~2.9]{SOZ92} and \cite[Section~9.3]{DEZ11}. Recall that a smooth compactly
supported vector field $\velo:\bbR^{3}\to\bbR^{3}$,
$\velo\in (C^{\infty}_{0}(\bbR^{3}))^{3}$, induces a one-parameter family
$\left(\Phibf_{s} =\Phibf(s;\cdot)\right)_{s\in\bbR}$ of diffeomorphisms of $\bbR^{3}$,
the associated \emph{flow} (map), defined as solution of the family of initial value
problems
\begin{gather}
  \label{eq:flow}
  \frac{\partial \Phibf}{\partial s}(s;\Bx) = \velo(\Phibf(s;\Bx))\;,\quad   \Phibf(0;\Bx) =
  \Bx \;,\quad \Bx\in\bbR^{3}\;,\quad s\in\bbR\;.
\end{gather}
For every parameter value ${s\in\bbR}$ the mapping ${\Bx\mapsto \Phi_{s}(\Bx)}$
induces a deformation of the domain $\Omega$ and of the medium, which impacts the static
electromagnetic fields. More precisely, for every parameter value $s\in\bbR$ we obtain a
unique field\footnote{Usually, forms defined on the transported domains $\Omega_{s}$ will be tagged
with $\wt{\mbox{ }}$.} $\wt{\omegabf}(s) := \extd \wt{\varphibf}(s) \in L^{2}\Lambda^{\ell+1}(\Omega_{s})$
defined on the transported domain $\Omega_{s}:=\Phibf_{s}(\Omega)$. The potentially
non-unique $s$-dependent potential $\wt{\varphibf}(s)\in\Hdf[\Omega_{s}]$, $s\in\bbR$, will solve
a version of \cref{eq:vffl} in a geometry and with a material deformed by $\Phibf_{s}$%
\skcomment{
My take on this:
\begin{itemize}
\item We are going to transport the domain - We also need to transport the data of the
  problem. This refers to the energy density, the material density, and the charge
  density.
\item They all transform with the respective pullbacks that are induced by the
  transport. This models what physicists would call "covariant" transformation.
\item The
  solution $\varphi$ does (in general) *not* transform by pullback, or does it? For each
  fixed s, the deformation can be seen as a (sophisticated) change of coordinates. With
  the pullback machinery, we just transform everything to the new coordinates. But this
  only retains the solution if all operators commute with (pullbacks of)
  diffeomorphisms. This is in general not the case for the material operator, it has a
  smaller symmetry group. This is somehow hidden in the innocent-looking $\Ce$. I have
  not fully understood this point yet - we should discuss it. 
\item I would expect that the
  energy density transforms like a 3-form, so why is the "dx" not affected by the
  transformation? If the "dx" is retained, it transforms like a 0-form (the "material
  point" assumption), which seems strange.
\item It might be that Sect. 4.1 sheds more light
  into this.
\end{itemize}
}:
\newcommand{\fent}{\wt{\Ce}_{\mathrm{loc}}}
\newcommand{\fentd}{\frac{\partial \fent}{\partial \omega}}
\begin{multline}
  \label{eq:vffs}
  \int\nolimits_{\Omega_{s}} \left< \frac{\partial \fent(s)}{\partial
      \omega}(\Bx,(\extd \wt{\varphibf}(s))(\Bx),\wt{\rhobf}(s)(\Bx)),\extd \wt{\etabf}
  \right> \\ = \int\nolimits_{\Omega_{s}}\wt{\sigmabf}(s)\wedge\wt{\etabf}\quad
  \forall \wt{\etabf} \in \Hdf[\Omega_{s}] \;,
\end{multline}
with a transported local field energy 
\begin{gather}
  \label{eq:fent}
  \fent(s)(\Bx,\omega,\rho) := \fen(\Phibf_{s}^{-1}(\Bx),\omega,\rho)\;,\quad \Bx\in
  \Omega_{s},\;
  \omega \in \Lambda^{\ell+1}(\bbR^{3}),\; \rho\in \Lambda^{3}(\bbR^{3})\;.
\end{gather}
Let us discuss the considerations underlying \cref{eq:vffs}:
\begin{enumerate}[label={(\roman{*})}]
\item In \cref{eq:fent} reflects the assumption that the field energy density $\fen$ is
  attached to ``material points'' \cite[Section~2]{BOS92}, \cite[IV,~(10)]{BOS11}, which
  is reflected by inserting $\Phibf_{s}^{-1}(\Bx)$ as its $\Bx$-argument.
\item Concerning the source field, we transport it with the flow,
  $\wt{\sigmabf}(s):=(\Phibf_{s}^{-1})^{*}\sigmabf$, where the ``$*$'' tags the
  \emph{pullback} of a differential form under transformation of domains
  \cite[Section~2.2.3]{HIP15}. Note that $\sigmabf(s)$ will still satisfy \cref{eq:ko}.
\item By conservation of mass, also the medium density is transformed by inverse pullback:
  $\wt{\rhobf}(s) = (\Phibf_{s}^{-1})^{*}\rhobf$, which is equivalent to demanding
  conservation of mass: $\int_{\Phibf_{s}(D)}\rhobf(s)=\int_{D}\rhobf$ for all ``control volumes''
  $D\subset \Omega$ and $s\in\bbR$.
\end{enumerate}

\begin{remark}
  \label{rem:ass}
  Item (ii) means that the source fields can be transported ``effortlessly'' and no energy
  is fed into or drained from the system in the process. This corresponds to the ``closed
  system property'' mentioned earlier.
\end{remark}

As already addressed in \cref{ss:hfc}, also the field energy will become $s$-dependent,
both, directly through the $s$-dependent domain of integration, the deformation of the
medium, and, indirectly, through the $s$-dependence of the electromagnetic field argument
$\wt{\varphibf}(s)\in L^{2}\Lambda^{\ell+1}(\Omega_{s})$. We end up with a field energy
function $\wt{\wen}:\bbR\to\bbR$,
\begin{gather}
  \label{eq:W}
  \wt{\wen}(s) := \int\nolimits_{\Omega_{s}}
  \fent(s)(\Bx,(\extd\wt{\varphibf}(s))(\Bx),\wt{\rhobf}(s)(\Bx))\,\mathrm{d}\Bx \;,
\end{gather}
where $\wt{\varphibf}(s)$ solves \cref{eq:vffs} for every $s\in\bbR$. 

Be aware that the flow $\left(\Phibf_{s}\right)_{s\in\bbR}$ is induced by the vector field
$\velo\in (C^{\infty}_{0}(\bbR^{3}))^{3}$, though this dependence is suppressed in the
notation. This makes $\wt{\wen}$ a function of $\velo$, as well, which suggests an
extended notation:
\begin{gather}
  \label{eq:1}
  \velo \quad \longrightarrow \quad \Phibf_{s}\quad \text{by \cref{eq:flow}}\quad
  \longrightarrow\quad s\mapsto \wt{\wen} = \wt{\wen}(\velo;s) \;.
\end{gather}
In light of these dependencies we define the \emph{shape} (Gateaux) \emph{derivative} of
the field energy in the direction of $\velo$,
\begin{gather}
  \label{eq:sgd}
  \left<\frac{d{\wen}}{d\Omega}(\Omega),\velo\right> := \lim\limits_{s\to 0}
  \frac{\wt{\wen}(\velo;s)-\wt{\wen}(\velo;0)}{s} \;,
\end{gather}
provided that the limit exists. In addition, if
$\velo\mapsto\left<\frac{d\wen}{d\Omega}(\Omega),\velo\right>$ is \emph{linear} and
\emph{continuous} on $ (C^{\infty}_{0}(\bbR^{3}))^{3}$ equipped with the topology of test
functions \cite[Chapter~6]{RUD73}, then the field energy $\wen$ is called \emph{shape
  differentiable} and its shape derivative $\frac{d\wen}{d\Omega}(\Omega)$ is a distribution on
the space of test vector fields, a distributional co-vector field.

Appealing to the virtual work principle we may call that distribution the \emph{force
  field} ${\phibf\in \left( (C^{\infty}_{0}(\bbR^{3}))^{3}\right)'}$ and state the
defining equation
\begin{gather}
  \label{eq:3}
  \wt{\wen}(\velo;s) - \wt{\wen}(\velo;0) = s \left<\phibf,\velo\right> + o(s)
  \quad\text{for}\quad s\to 0\;,\quad \forall \velo \in (C^{\infty}_{0}(\bbR^{3}))^{3} \;,
\end{gather}
for which a shorthand notation could be
\begin{gather}
  \label{eq:2}
  \boxed{\phibf := \frac{d{\wen}}{d\Omega}(\Omega)}
  \quad\overset{\text{\cref{eq:sgd}}}{\Rightarrow}\quad
  \left<\phibf,\velo\right> =  \lim\limits_{s\to 0}
  \frac{\wt{\wen}(\velo;s)-\wt{\wen}(\velo;0)}{s}\;.
\end{gather}
If $\phibf$ is continuous even on the Sobolev space $(\Ltwo[\bbR^{3}])^{3}$, then
\skcomment{Otherwise it could (possibly) be seen as a de Rham current.}
we can
regard $\phibf$ as a 1-form ${\in L^{2}\Lambda^{1}(\bbR^{3})}$ and its Euclidean
vector proxy is a vector field ${\Vf\in(\Ltwo[\bbR^{3}])^{3} }$, which satisfies
\begin{gather}
  \label{eq:17}
  \int\nolimits_{\Omega}\Vf(\Bx)\cdot \velo(\Bx)\,\mathrm{d}\Bx =
  \left<\frac{d{\wen}}{d\Omega}(\Omega),\velo\right>\quad\forall \velo \in
   (C^{\infty}_{0}(\bbR^{3}))^{3} \;,
\end{gather}
harking back to \cref{eq:40}, \emph{cf.} the interpretation of force as a covector-valued
3-form in \cite[Section~I]{BOS90f}, \cite[Section~4.2]{BOS92}, and
\cite[Section~III]{BOS05}.

\section{Force fields by the adjoint method of shape differentiation}
\label{sec:force}

Shape differentiating $\wt{\wen}$ amounts to shape differentiating a functional that
depends on both the deformation (expressed through the argument~$s$) and an $s$-dependent
field, which solves the variational problem \cref{eq:vffs}: we face the task of shape
differentiation under a variational constraint.

\subsection{Transformation to reference domain}
\label{ss:trf}

For our envisaged mathematical approach it is essential to remove the {$s$-dependence}
from the trial and test space ${\Hdf[\Omega_{s}]}$ in \cref{eq:vffs}. This can be achieved
by pulling the variational equation \eqref{eq:vffs} and the energy functional $\wt{\wen}$
from \cref{eq:W} back to the ``reference domain'' $\Omega$ \cite[Section~2.11]{SOZ92},
also called ``material manifold'' in \cite{BOS92}. Using the transformation formula for
multi-dimensional integrals and inverse pullbacks of differential forms via
$\Phibf_{s}^{-1}:\Omega_{s}\to \Omega$, \eqref{eq:vffs} can be converted into: For any
$s\in\bbR$ seek $\wh{\varphibf}(s):=\Phibf_{s}^{*}\wt{\varphibf}(s)\in\Hdf$ such that
\newcommand{\pbfent}{\left(\Phibf_{s}^{*}\frac{\partial \fent(s)}{\partial \omega}\right)}
\begin{multline}
  \label{eq:vfft}
  \int\nolimits_{\Omega}
    \Phibf_{s}^{*}\Bigg\{\Bx\mapsto  \Bigl<
    \frac{\partial \fent(s)}{\partial \omega}\left(\Bx,\left(\extd
      (\Phibf_{s}^{-1})^{*}\wh{\varphibf}(s)\right)(\Bx),\wt{\rhobf}(\Bx)\right),
    (\extd(\Phibf_{s}^{-1})^{*}{\etabf})(\Bx)
  \Bigr>\Biggr\}  \\ = \int\nolimits_{\Omega}\sigmabf\wedge{\etabf}\quad\forall \etabf \in \Hdf \;.
\end{multline}
In the sequel, with a $\widehat{\mbox{ }}$\ we label forms defined on $\Omega$ and related to forms on
$\Omega_{s}$ according to $\wt{\nubf}(s) = (\Phibf_{s}^{-1})^{*}\wh{\nubf}$, $\wh{\nubf}$
a $k$-form. 
As regards the right-hand side in \cref{eq:vfft} we have exploited the invariance of
integrals under pullback,
\begin{gather}
  \label{eq:19}
  \int\nolimits_{\Omega}\sigmabf\wedge {\etabf} = 
  \int\nolimits_{\Omega}\Phibf_{s}^{*}\wt{\sigmabf}(s)\wedge {\etabf} =
  \int\nolimits_{\Omega_{s}}\wt{\sigmabf}(s)\wedge (\Phibf_{s}^{-1})^{*}{\etabf} \;.
\end{gather}
We remind, that pullback and exterior derivative commute, which ensures that 
\cref{eq:vfft} remains a variational equation with a consistent right-hand side. 

The same transformation to the reference domain can be applied to the field energy $\wt{\wen}$ 
and yields $\wt{\wen}(\velo;s) = \wh{\wen}(s,\wh{\varphibf}(s),\rhobf)$ with 
\begin{multline}
  \label{eq:5}
  \wh{\wen}(s,{\etabf},\rhobf) := \int\nolimits_{\Omega}
  \Phibf_{s}^{*}\left\{\Bx\mapsto\fent(s)(\Bx,(\extd(\Phibf_{s}^{-1})^{*}{\etabf})(\Bx),
  \wt{\rhobf}(\Bx))\right\},\; {\etabf}\in \Hdf\;.
\end{multline}
We observe that the structure of \cref{eq:vff} carries over and that \eqref{eq:vfft} can
be written as
\begin{gather}
  \label{eq:vfftd}
  \wh{\varphibf}(s)\in\Hdf:\quad
  \left<
    \frac{\partial \wh{\wen}}{\partial {\etabf}}(s;\wh{\varphibf}(s),\rhobf),{\etabf}
  \right> =  \int\nolimits_{\Omega}\sigmabf\wedge{\etabf}\quad\forall {\etabf} \in \Hdf \;.
\end{gather}

\subsection{Adjoint method: abstract view}
\label{ss:adjm}

We briefly review a fundamental technique, known as adjoint approach
\cite[Sect.~1.6.2]{HPU09}, \cite[Sect.~2.3.3]{GUN03}, \cite[\S2.10]{TRO10}, devised for
the differentiation of functionals under variational constraints. Since it provides the
key ideas for this article, we are going to discuss it in detail, and this is best done
from an abstract perspective.

On a Banach space $H$ we consider an ``energy'' functional $F:\bbR\times H\to \bbR$,
$F=F(s,\Vw)$, which is differentiable (with Lipschitz continuous derivative) and strictly
convex in the second argument, manifesting itself in the fact that
$\frac{\partial F}{\partial \Vw}(s,\cdot):H\to H'$, $H'$ the dual space of $H$, is
$s$-uniformly strictly monotone. In addition, $F$ must be continuously differentiable with
respect to $s$.

We are interested in the reduced functional $\wt{F}:\bbR\to\bbR$, $\wt{F}(s) :=
F(s,\Vu(s))$, where $\Vu(s)\in H$ solves the \emph{state variational problem}
\begin{gather}
  \label{eq:6}
  \Vu(s)\in H:\quad \left<\frac{\partial F}{\partial \Vw}(s,\Vu(s)),\Vv\right> =
  \left<\ell(s),\Vv\right>\quad\forall \Vv\in H \;,
\end{gather}
where $\ell:\bbR\to H'$ is a smooth family of linear functionals. The assumptions on $F$
guarantee existence and uniqueness of solutions. 

Our goal is to compute the derivative $\frac{d\wt{F}}{ds}(0)$ of $\wt{F}$ in $s=0$. To
that end we introduce the \emph{Lagrangian functional}
\begin{gather}
  \label{eq:LF}
  \Cl:
  \left\{
    \begin{array}[c]{ccl}
      \bbR\times H \times H &\to& \bbR \;,\\
      (s,\Vw,\Vv) &\mapsto & F(s,\Vw) +
      \left<\dfrac{\partial F}{\partial \Vw}(s,\Vw),\Vv
      \right> - \left<\ell(s),\Vv\right>\;.
      \end{array}
  \right.
\end{gather}
From \eqref{eq:6} we conclude that 
\begin{gather}
  \label{eq:Lt}
  \wt{F}(s) = \Cl(s,\Vu(s),\Vv) \quad \text{\bf for any}\quad \Vv\in H \;.
\end{gather}
Setting $\Vv$ to some suitable fixed \emph{$s$-independent} element of $H$, we infer by the
chain rule
\begin{gather}
  \label{eq:4}
  \frac{d\wt{F}}{ds}(s) = \frac{\partial \Cl}{\partial s}(s,\Vu(s),\Vv) +
  \left<\frac{\partial\Cl}{\partial \Vw}(s,\Vu(s),\Vv),\frac{d\Vu}{ds}(s) \right>\;.
\end{gather}
The key idea is to choose $\Vv$ as the solution of the \emph{adjoint variational problem}
\begin{gather}
  \label{eq:7}
  \Vp\in H:\quad
  \left<
    \frac{\partial\Cl}{\partial\Vw}(0,\Vu(0),\Vp),\Vq
  \right>  = 0 \quad\forall \Vq \in H \;,
\end{gather}
such that the second term in \cref{eq:4} containing the elusive derivative
$\frac{d\Vu}{ds}$ can be dropped. From the definition of $\Cl$ we derive a more detailed
expression for $\frac{\partial\Cl}{\partial\Vw}$, 
\begin{gather}
  \label{eq:8}
  \left<
    \frac{\partial \Cl}{\partial \Vw}(s,\Vw,\Vv),\Vq
  \right>  =
  \left<
    \frac{\partial F}{\partial \Vw}(s,\Vw),\Vq
  \right> + \left<
    \frac{\partial^{2}F}{\partial\Vw^{2}}(s,\Vw),(\Vv,\Vq)
  \right> \;.
\end{gather}
Thus \cref{eq:7} becomes the linear \emph{adjoint variational problem}, 
\begin{gather}
  \label{eq:9}
  \Vp\in H:\quad
  \left<
    \frac{\partial^{2}F}{\partial\Vw^{2}}(0,\Vu(0)),(\Vp,\Vq)
  \right> =  -\left<
    \frac{\partial F}{\partial \Vw}(0,\Vu(0)),\Vq
  \right>\quad\forall \Vq\in H \;.
\end{gather}
In light of \cref{eq:4} and \cref{eq:7} we conclude, $\Vp\in H$ the solution of
\cref{eq:9},
\begin{gather}
  \label{eq:10}
  \frac{d\wt{F}}{ds}(0) = \frac{\partial \Cl}{\partial s}(0,\Vu(0),\Vp) \;,
\end{gather}
which, using \cref{eq:LF}, reads more concretely
\begin{gather}
  \label{eq:11}
  \frac{d\wt{F}}{ds}(0) = \frac{\partial F}{\partial s}(0,\Vu(0)) +
  \left<
    \frac{\partial^{2}F}{\partial s\partial \Vw}(0,\Vu(0)),\Vp
  \right> -
  \left<
    \frac{d\ell}{ds}(0),\Vp
  \right> \;.
\end{gather}
Note that all derivatives with respect to $s$ are partial derivatives! We can, thus, pull
them in front and we arrive at a \emph{total $s$-derivative},
\begin{gather}
  \label{eq:12}
  \boxed{\frac{d\wt{F}}{ds}(0) = \frac{d}{ds} {\left\{
    F(s,\Vu(0)) + \left<\frac{\partial F}{\partial \Vw}(s,\Vu(0)),\Vp \right> -
    \left<\ell(s),\Vp\right>
  \right\}_{|s=0}}}\;.
\end{gather}
Though it is clear from the formula, we emphasize that neither the so-called state
solution $\Vu(0)$ nor the adjoint solution $\Vp$ depend on $s$. 

\subsection{Shape derivative of field energy}
\label{ss:sdfe}

We apply the abstract theory of \Cref{ss:adjm} to \cref{eq:5} with $H \leftrightarrow \Hdf$,
\begin{align*}
  F(s,\Vw)\quad \leftrightarrow\quad && (s,\wh{\etabf}) & \mapsto
  \wh{\wen}(s,\wh{\etabf},\rhobf)\;, && \rhobf\quad\text{a parameter}\;,\\
  \wt{F}(s) \quad \leftrightarrow\quad&& s & \mapsto
  \wt{\wen}(\velo;s) = \wh{\wen}(s,\wh{\varphibf}(s),\rhobf)\;,&& \velo\quad\text{a parameter}\;.
\end{align*}
Then the role of the state equation \cref{eq:6} is played by \cref{eq:vfft} and
$\wh{\varphibf}(s)$ has to be substituted for the state solution $\Vu(s)$. Applying the
chain rule, we find that the abstract adjoint variational problem \cref{eq:9} becomes the
following linear variational problem: seek ${\pibf\in \Hdf}$ such that
\begin{multline}
  \label{eq:eavp}
  \int\nolimits_{\Omega}
  \left<\frac{\partial^{2}\fen}{\partial \omega^{2}}(\rfx,(\extd
    \wh{\varphibf}(0))(\rfx),\rhobf(\rfx)),
    \left((\extd \pibf)(\rfx),(\extd\nubf)(\rfx)\right)
  \right> \\
  = - \int\nolimits_{\Omega}
  \left<
    \frac{\partial \fen}{\partial
      \omega}(\rfx,(\extd\wh{\varphibf}(0))(\rfx),\rhobf(\rfx)),(\extd \nubf)(\rfx)
  \right> \quad\forall \nubf\in\Hdf \;.
\end{multline}
Obviously, the right-hand side linear functional evaluates to zero, if $\nubf$ belongs to
the kernel of $\extd$. This means that also \cref{eq:eavp} features a consistent
right-hand side. Thus \cref{eq:eavp} is a valid adjoint variational problem \cref{eq:6},
though we assumed existence and uniqueness of solutions in the abstract framework;
switching to the orthogonal complement of the kernel of $\extd$ will remedy
non-uniqueness.
   
With the ($s$-independent!) state solution
${\varphibf}:=\wh{\varphibf}(0)$ and the adjoint solution
$\pibf$ at our disposal, we can now state the concrete incarnation of \cref{eq:12}.
\begin{multline*}
  \frac{\partial\wt{\wen}}{\partial s}(\velo;0)
  = \frac{d}{ds} \left\{ \wh{\wen}(s,\varphibf,\rhobf)
      + \left<\frac{\partial \wh{\wen}}{\partial \omegabf}(s,\varphibf,\rhobf),\pibf
      \right> - \int\nolimits_{\Omega}\sigmabf\wedge\pibf
    \right\}_{|s=0} \\
  =
\begin{aligned}[t]
  \frac{d}{ds}
  \Biggl\{ &
  \int\nolimits_{\Omega}\Phibf_{s}^{*}\left\{\Bx\mapsto
    \fent(s)(\Bx,(\extd (\Phibf_{s}^{-1})^{*}\varphibf)(\Bx),
  \wt{\rhobf}(\Bx))\right\}
  \;+ 
  \\
  & \int\nolimits_{\Omega}
  \begin{aligned}[t]
    \Phibf_{s}^{*}\Biggl\{\Bx\mapsto \Bigl< \frac{\partial \fent(s)}{\partial \omega}(\Bx,(\extd
       (\Phibf_{s}^{-1})^{*}\varphibf)(\Bx),\wt{\rhobf}(\Bx)),(\extd
      (\Phibf_{s}^{-1})^{*}\pibf)(\Bx) \Bigr>\Biggr\}
      - 
    \end{aligned} \\
    & 
    \cancelto{= \text{const}}{\int\nolimits_{\Omega}\sigmabf\wedge\pibf}
    \phantom{= \text{const}}\quad \Biggr\}_{|s=0} \;.
\end{aligned}
\end{multline*}
Next, we
\begin{enumerate}[label={(\roman{*})}]
\item undo the transformation to the reference domain $\Omega$ and return to integrals over
  $\Omega_{s}$, and 
\item apply the local material law \cref{eq:lml}, $\lml(\Bx,\omega,\rho)\wedge \nu =
  \left<\frac{\partial \fen}{\partial \omega}(\Bx,\omega,\rho),\nu\right>$.
\end{enumerate}
This yields
\begin{gather}
  \label{eq:star}
  \frac{\partial\wt{\wen}}{\partial s}(\velo;0) = 
  \begin{aligned}[t]
  \frac{d}{ds}
  \Biggl\{ &
  \int\nolimits_{\Omega_{s}}\fen(\Phibf_{s}^{-1}(\Bx),
  (\extd (\Phibf_{s}^{-1})^{*}\varphibf)(\Bx),
  ((\Phibf_{s}^{-1})^{*}\rhobf)(\Bx))  + 
  \\
  & \int\nolimits_{\Omega_{s}}
  \begin{aligned}[t]
    \lml(\Phibf_{s}^{-1}(\Bx), (\extd (\Phibf_{s}^{-1})^{*}\varphibf)(\Bx),
    ((\Phibf_{s}^{-1})^{*}\rhobf)(\Bx)) \wedge \\
    (\extd (\Phibf_{s}^{-1})^{*}\pibf)(\Bx)
    \Biggr\}_{|s=0} \;.
  \end{aligned}
\end{aligned}
\end{gather}
Let us introduce the \emph{frozen fields}\footnote{For a fixed $s\in\bbR$ the frozen field
$\overline{\nubf}$ is defined on $\Omega_{s}$, which is not entirely consistent with
notational conventions, because forms on $\Omega_{s}$ should wear a $\wt{\mbox{ }}$.}
\begin{gather}
  \label{eq:13}
  \froz{\varphibf}(s) := (\Phibf_{s}^{-1})^{*}\varphibf\;,\quad
  \froz{\pibf}(s) := (\Phibf_{s}^{-1})^{*}\pibf\;,\quad
  \froz{\rhobf}(s) := (\Phibf_{s}^{-1})^{*}\rhobf\;.
\end{gather}
They owe their name to the property that, with $\nubf$ standing for either
$\varphibf$, $\pibf$, or $\rhobf$, and $S\subset \Omega$ for any oriented
$k$-dimensional surface, $k\in \{0,1,2,3\}$ being the degree of the form $\nubf$,
\begin{gather}
  \label{eq:15}
  \frac{d}{ds}\int\limits_{\Phibf_{s}(S)} \froz{\nubf}(s) = 0 \quad
  \Longleftrightarrow\quad
  \boxed{\frac{D\froz{\nubf}}{D\velo} = \frac{d\froz{\nubf}}{ds} + \Lie\froz{\nubf} = 0} \;, 
\end{gather}
where
\begin{itemize}
\item $\frac{D}{D\velo}$ is the \emph{material derivative operator} with respect to the
  vector field $\velo$ \cite[Definition~1.1]{HIL17}, and
\item $\Lie$ is the \emph{Lie derivative operator} induced by the velocity field $\velo$
  \cite[V,~\S5]{LAN99}.
\end{itemize}
In other words, rephrasing \cite[Section~II]{BOS05},
\begin{itemize}
\item the frozen fields are just \emph{passively advected} with the flow $\Phibf$, 
\item they are fields, whose ``fluxes are held constant'' while carried along by the
  velocity field $\velo$, and
\item their integrals on test surfaces moving with the flow are constant with respect to
  variations of $s$, as we have already remarked in \Cref{ss:hfc}.
\end{itemize}

Relying on the frozen fields, we can rewrite the above formula \cref{eq:star} as the following expression
for the force field, \emph{cf.} \cref{eq:2},
\begin{gather}
  \label{eq:main}
  \boxed{
    \left<\phibf,\velo\right> = 
      \begin{aligned}[t]
        \frac{d}{ds}
        \Biggl\{ &
        \int\nolimits_{\Omega_{s}}\fen(\Phibf_{s}^{-1}(\Bx),
        (\extd \froz{\varphibf}(s))(\Bx),\froz{\rhobf}(s)(\Bx))  + 
        \\
        & \int\nolimits_{\Omega_{s}}
        \begin{aligned}[t]
          \lml(\Phibf_{s}^{-1}(\Bx), (\extd \froz{\varphibf}(s))(\Bx),
          \froz{\rhobf}(s)(\Bx)) \wedge
          (\extd \froz{\pibf}(s))(\Bx)
          \Biggr\}_{|s=0} .
        \end{aligned}
    \end{aligned}}
\end{gather}
In plain English, 
\begin{center}
  \mycolorbox{0.9\linewidth}{%
    the distributional force 1-form evaluated for $\velo$, is the $s$-derivative of
    energy-like expressions on $\velo$-transported media, but just supplied with frozen
    $s$-dependent fields.  }
\end{center}
\medskip

The formula \cref{eq:main} can be manipulated further using a central formula of shape
calculus
\cite[Theorem~1]{HIL11}
\begin{gather}
  \label{eq:16}
  \left.\frac{d}{ds}\int\nolimits_{\Omega_{s}}\nubf(s)\right|_{s=0} =
  \int\nolimits_{\partial\Omega} \contr \nubf(0) +
  \int\nolimits_{\Omega}\frac{d \nubf}{d s}(0)
\end{gather}
for any parameter-dependent 3-form $\nubf=\nubf(s)$:
\newcommand{\xarg}{\Bx,(\extd\varphibf)(\Bx),\rhobf(\Bx)}
\newcommand{\sarg}{\Phibf_{s}^{-1}(\Bx),(\extd \froz{\varphibf}(s))(\Bx),\froz{\rhobf}(s)(\Bx)}
\begin{gather}
  \label{eq:37}
  \left<\phibf,\velo\right> =
  \begin{aligned}[t]
    & \int\nolimits_{\partial\Omega}\contr \left\{\Bx\mapsto \fen(\xarg)\right\} + \\
    & \int\nolimits_{\Omega}\frac{d}{ds} \left\{\fen(\sarg)\right\}_{|s=0} + \\
    & \int\nolimits_{\partial\Omega}
    \contr \left\{\Bx\mapsto
      \lml(\xarg)\wedge\extd \pibf(\Bx)\right\}
    + \\
    & \int\nolimits_{\Omega} \frac{d}{ds}\left\{
      \lml(\sarg)\wedge \extd \froz{\pibf}(s)(\Bx)
    \right\}_{|s=0} \;,
  \end{aligned}
\end{gather}
where we exploited that ${\Phibf_{0}=\Id}$. The total $s$-derivatives under the
integrals can be expanded further based on the formula for the $s$-derivative of frozen fields,
\begin{gather}
  \label{eq:Lie}
  \frac{d}{ds}\left\{s\mapsto \left((\Phibf_{s}^{-1})^{*}\nubf\right)\right\}_{|s=0} = -
  (\Lie \nubf)(\Bx)\;,\quad \Bx\in \Omega\;.
\end{gather}
We also take into account
$\frac{d}{ds}\left\{\Phibf_{s}^{-1}(\Bx)\right\}_{|s=0}=-\velo(\Bx)$, ${\Bx\in\Omega}$,
and get
\begin{multline}
  \label{eq:36}
  \frac{d}{ds} \left\{\fen(\sarg)\right\}_{|s=0} = \\
  \begin{aligned}[t]
    - \Bigl( &  \left<\frac{\partial\fen}{\partial \Bx}(\xarg),\velo(\Bx) \right> + \\
    & \cob{\left<\frac{\partial \fen}{\partial \omega}(\xarg),(\Lie \extd\varphibf)(\Bx)\right>} + \\
    & \left<\frac{\partial \fen}{\partial \rho}(\xarg),(\Lie\rhobf)(\Bx)\right>\Bigr) \;,
  \end{aligned}
\end{multline}
\begin{multline}
  \label{eq:38}
  \frac{d}{ds}\left\{ \lml(\sarg)\wedge (\extd
    \froz{\pibf}(s))(\Bx) \right\}_{|s=0} \\
   = \begin{aligned}[t]
     & \frac{d}{ds}\left\{\lml(\sarg)\right\}_{|s=0} \wedge (\extd \pibf)(\Bx) - \\ & 
     \quad \lml(\xarg) \wedge (\Lie \extd \pibf)(\Bx)
   \end{aligned} \\
   =
   \begin{aligned}[t]
     -\Bigl(
     \begin{aligned}[t]
       & \left<
         \frac{\partial \lml}{\partial \Bx}(\xarg),\velo(\Bx)
       \right> + \\
       & \left<
         \frac{\partial\lml}{\partial \omega}(\xarg), (\Lie\extd\varphibf)(\Bx)
       \right> + \\
       & \left<
         \frac{\partial \lml}{\partial \rho}(\xarg), (\Lie\rhobf)(\Bx)
       \right>      \Bigr) \wedge \extd\pibf(\Bx)  -
     \end{aligned}   \\
     \cob{\lml(\xarg) \wedge (\Lie \extd \pibf)(\Bx)} \;.
   \end{aligned}
\end{multline}
From these formulas we can conclude that the expressions we got for
$\left<\phibf,\velo\right>$ are linear in $\velo$ indeed, because $\velo\mapsto
\Lie\nubf$ is linear. We also conclude that the right-hand sides in \cref{eq:36} and
\cref{eq:38} are continuous on $(C^{\infty}_{0}(\bbR^{3}))^{3}$ provided that $\fen$
enjoys some very weak regularity. Summing up, \cref{eq:main}/\cref{eq:37} define a
distribution on $(C^{\infty}_{0}(\bbR^{3}))^{3}$.

Notice that the contributions of the \cob{blue} terms in \cref{eq:36} and \cref{eq:38} can
be simplified based on \cref{eq:vffl}/\cref{eq:vfflm} and the fundamental homotopy
formula \cite[Proposition~5.3]{LAN99}
\begin{gather}
  \label{eq:HF}
  \Lie = \extd\circ \contr + \contr \circ \extd \quad \Rightarrow \quad
  \Lie\circ\extd = \extd\circ\contr \circ \extd \;, 
\end{gather}
where $\contr$ denotes the \emph{contraction} of a form with the vector field $\velo$
\cite[V \S 5]{LAN99}, 
\begin{align}
  \label{eq:44}
  \int\nolimits_{\Omega}
  \cob{\left<\frac{\partial \fen}{\partial \omega}(\xarg),(\Lie
      \extd\varphibf)(\Bx)\right>}
  & = \int\nolimits_{\Omega}\sigmabf\wedge \contr (\extd \varphibf) \;, \\
  \label{eq:45}
  \int\nolimits_{\Omega}
  \cob{\lml(\xarg) \wedge (\Lie \extd \pibf)(\Bx)} & =
  \int\nolimits_{\Omega} \sigmabf\wedge  \contr (\extd \pibf) \;. 
\end{align}
This is possible, if $\Lie\varphibf,\Lie\pibf\in\Hdf$.

\begin{remark}
  \label{rem:jump}
  If either $\fen$ or $\lml$, regarded as functions of $\Bx$, feature a discontinuity
  across a closed surface $\Sigma\subset\Omega$, then both $\frac{\partial \fen}{\partial\Bx}$
  and $\frac{\partial\lml}{\partial \Bx}$ have to be viewed as distributional derivatives.
  For their integrals this means, e.g.,
  \begin{multline}
    \label{eq:39}
    \int\nolimits_{\Omega}\left<\frac{\partial\fen}{\partial\Bx}(\Bx,\ldots),\velo(\Bx)\right>
    \,\mathrm{d}\Bx  \\ =
    \int\nolimits_{\Omega\setminus\Sigma}
    \left<\frac{\partial\fen}{\partial\Bx}(\Bx,\ldots),\velo(\Bx)\right>
    \,\mathrm{d}\Bx +
    \int\nolimits_{\Sigma}
    \left[\fen(\cdot,\ldots)\right]_{\Sigma}(\Bx)(\velo(\Bx)\cdot\Bn_{\Sigma}(\Bx))\,
    \mathrm{d}S(\Bx) \;,
  \end{multline}
  where $\left[\cdot\right]_{\Sigma}$ designates the height of the jump across $\Sigma$
  and $\Bn_{\Sigma}$ is a unit normal vector field on $\Sigma$, its direction matching
  $\left[\cdot\right]_{\Sigma}$. 
\end{remark}

\subsection{Vector proxy incarnations of force formulas}
\label{ss:vpff}

As in \cite[Section~.2.2.2]{HIP15} we use an overset arrow to indicate the identification
of a differential $k$-form with its Euclidean vector proxy in $\bbR^{3}$, which is a
vector field with $\binom{3}{k}$ components. From \cite[Table~2]{HHP15},
\cite[Table~2.1]{AFW06}, we cite the classical vector proxy incarnations of various
operators for a smooth differential $k$-forms $\etabf$:
\begin{subequations}
  \label{eq:30}
  \begin{align}
    \label{eq:30a}
    &k=0: & \vpr{\extd \etabf} & = \grad \vpr{\etabf} \;, & \vpr{\contr\etabf} & = 0\;, \\
    \label{eq:30b}
    &k=1: & \vpr{\extd\etabf} & = \curl\vpr{\etabf}\;, & \vpr{\contr\etabf} & =
    \velo\cdot\vpr{\etabf} \;, \\
    \label{eq:30c}
    &k=2: & \vpr{\extd\etabf} & = \Div\vpr{\etabf}\;, & \vpr{\contr\etabf} &
    = \velo\times \vpr{\etabf}\;, \\
    \label{eq:30d}
    &k=3: & \vpr{\extd\etabf} & = 0 \;, & \vpr{\contr\etabf} & =  \velo\,\vpr{\etabf}\;.
  \end{align}
\end{subequations}
The Lie derivative in vector proxies can then be deduced from \cref{eq:HF} \cite[Table~4]{HIL11}.

Based on \cref{eq:30} we translate the derived force formulas into the vector calculus
language of classical electromagnetism. In the sequel we use the notation $r:\Omega\to \bbR$ for
the vector proxy $\vpr{\rhobf}$ of the density 3-form $\rhobf$.

\renewcommand{\zbHone}{H^{1}_{0,w}(\Omega)}
\renewcommand{\zbHcurl}{\BH_{0,w}(\curl,\Omega)}

\subsubsection{Electrostatics}
\label{sss:e}

This is the case $\ell=0$. The local energy density $\fen$ is a function
$\Omega\times\bbR^{3}\times\bbR\to\bbR^{+}$, $\fen=\fen(\Bx,\Vw,r)$, and we have the correspondences
\begin{gather}
  \label{eq:31}
  V = \vpr{\varphibf}\;,\quad \VE = -\grad V = -\vpr{\extd\varphibf}\;,\\
  \label{eq:32}
  \vpr{\lml(\Bx,\omega,\rho)} = -\VD(\Bx,-\Vw,r)\;,\quad \Vw := \vpr{\omega},\; r:=\vpr{\rho} \;. 
\end{gather}
Remember that $\VD$ is the displacement field due to the electric field $\VE$ and both are
linked through a local material law \cref{eq:mlloc}. Now we can state the vector-proxy
version of the frozen-field force formula \eqref{eq:main}
\begin{multline}
  \label{eq:34}
  \int\nolimits_{\Omega}\Vf(\Bx)\cdot\velo(\Bx)\,\mathrm{d}\Bx \\ = 
  \frac{d}{ds} \Bigl\{
  \begin{aligned}[t]
    \int\nolimits_{\Omega_{s}}
    & \fen(\Phibf_{s}^{-1}(\Bx),(\grad\overline{V}(s))(\Bx),\overline{r}(s)(\Bx)) - \\ &
    \VD(\Phibf_{s}^{-1}(\Bx),(-\grad\overline{V}(s))(\Bx),\overline{r}(s)(\Bx))\cdot
    (\grad\overline{P}(s))(\Bx)\,\mathrm{d}\Bx\Bigr\}_{|s=0} \;.
  \end{aligned}
\end{multline}
Here $P\in\zbHone$ is the solution of the adjoint variational equation
\begin{multline}
  \label{eq:eavpel}
  \int\nolimits_{\Omega}
  \frac{\partial^{2}\fen}{\partial \Vw^{2}}(\Bx,-\VE(\Bx),r(\Bx))
    \grad P(\Bx)\cdot\grad W(\Bx)
    \,\mathrm{d}\Bx \\
    =  \int\nolimits_{\Omega}
    \VD(\Bx,\VE(\Bx),r(\Bx))\cdot\grad W(\Bx)\,\mathrm{d}\Bx\quad\forall W \in\zbHone\;,
\end{multline}
and the frozen fields have to be understood as
\begin{gather}
  \label{eq:35}
  \overline{V}(s;\Bx) = V(\rfx)\;,\quad
  \overline{P}(s;\Bx) = P(\rfx)\;,\quad
  \overline{r}(s;\Bx) = |\det\Derv\Phibf_{s}(\rfx)|^{-1}r(\rfx)\;,
\end{gather}
for $\Bx\in\Omega_{s}$, $\rfx := \Phibf_{s}^{-1}(\Bx)$.
The force formula \cref{eq:37}+\cref{eq:36}+\cref{eq:38} in vector-proxy notation turns out as
\begin{multline}
  \label{eq:33}
  \int\nolimits_{\Omega}\Vf(\Bx)\cdot\velo(\Bx)\,\mathrm{d}\Bx \\ =
  \begin{aligned}[t]
    & \int\nolimits_{\partial\Omega}\fen(\Bx,-\VE(\Bx),r(\Bx))(\velo(\Bx)\cdot\Bn(\Bx))\,\mathrm{d}S(\Bx)
    - \\ & \int\nolimits_{\Omega}
    \begin{aligned}[t]
      \Bigl\{ & \grad_{\Bx}\fen(\Bx,-\VE(\Bx)),r(\Bx))\cdot \velo(\Bx)  -       \\ 
      & \grad_{\Vw}\fen(\Bx,-\VE(\Bx)),r(\Bx))\cdot \grad(\velo\cdot\VE)(\Bx) + \\
      & \frac{\partial \fen}{\partial r}(\Bx,-\VE(\Bx)),r(\Bx))\,\Div(r\,\velo)(\Bx)\Bigr\}
      \,\mathrm{d}\Bx -
    \end{aligned} \\
    & \int\nolimits_{\partial\Omega}(\VD(\Bx,\VE(\Bx),r(\Bx))\cdot \grad P(\Bx))
    (\velo(\Bx)\cdot\Bn(\Bx))\,\mathrm{d}S(\Bx)
    + \\ & \int\nolimits_{\Omega}
    \begin{aligned}[t]
      \Bigl\{\Bigl(& \grad_{\Bx}\VD(\Bx,\VE(\Bx),r(\Bx))\cdot\velo(\Bx) - \\ &
      \grad_{\Vw}\VD(\Bx,\VE(\Bx),r(\Bx))\cdot \grad(\velo\cdot\VE)(\Bx) + \\ &
      \frac{\partial\VD}{\partial r}(\Bx,\VE(\Bx),r(\Bx))\, \Div(r\,\velo)(\Bx)\Bigr)\cdot
      \grad P(\Bx)  + \\ &
      \VD(\Bx,\VE(\Bx),r(\Bx))\cdot \grad(\velo\cdot\grad P)(\Bx)\Bigr\}\,\mathrm{d}\Bx
    \end{aligned}
  \end{aligned}
\end{multline}

\subsubsection{Magnetostatics}
\label{sss:m}

Now we consider $\ell=1$, $\varphibf$ is a 1-form. Again, the local energy density $\fen$
is a function $\Omega\times\bbR^{3}\times\bbR\to\bbR^{+}$, $\fen=\fen(\Bx,\Vb,r)$, and the
relevant correspondences are
\begin{gather}
  \label{eq:31m}
  \VA = \vpr{\varphibf}\;,\quad \VB = \curl\VA = \vpr{\extd\varphibf}\;,\\
  \label{eq:32m}
  \vpr{\lml(\Bx,\omega,\rho)} = \VH(\Bx,\Vb,r)\;,\quad \Vb := \vpr{\omega},\; r:=\vpr{\rho} \;. 
\end{gather}
Now the frozen-field force formula reads
\begin{multline}
  \label{eq:34m}
  \int\nolimits_{\Omega}\Vf(\Bx)\cdot\velo(\Bx)\,\mathrm{d}\Bx \\ = 
  \frac{d}{ds} \Bigl\{
  \begin{aligned}[t]
    \int\nolimits_{\Omega_{s}}
    & \fen(\Phibf_{s}^{-1}(\Bx),(\curl\overline{\VA}(s))(\Bx),\overline{r}(s)(\Bx)) + \\ &
    \VH(\Phibf_{s}^{-1}(\Bx),(\curl\overline{\VA}(s))(\Bx),\overline{r}(s)(\Bx))\cdot
    (\curl\overline{\VP}(s))(\Bx)\,\mathrm{d}\Bx\Bigr\}_{|s=0} \;.
  \end{aligned}
\end{multline}
The adjoint solution $\VP\in\zbHcurl$ is defined as solution of the adjoint variational
equation
\begin{multline}
  \label{eq:eavpm}
  \int\nolimits_{\Omega}
  \frac{\partial^{2}\fen}{\partial \Vb^{2}}(\Bx,\VB(\Bx),r(\Bx))
    \curl \VP(\Bx)\cdot\curl \VW(\Bx)
    \,\mathrm{d}\rfx \\
    =  -\int\nolimits_{\Omega}
    \VH(\Bx,\VB(\Bx),r(\Bx))\cdot\curl \VW(\Bx)\,\mathrm{d}\rfx\quad\forall \VW\in\zbHcurl \;,
\end{multline}
and freezing the fields now means
\begin{gather}
  \label{eq:35m}
  \begin{aligned}
    \overline{\VA}(s;\Bx) & = (\Derv\Phibf_{s}(\rfx))^{-\top}\VA(\rfx)\;,\\
    \overline{\VP}(s;\Bx) & = (\Derv\Phibf_{s}(\rfx))^{-\top} \VP(\rfx)\;,\\
    \overline{r}(s;\Bx) & = |\det\Derv\Phibf_{s}(\rfx)|^{-1}r(\rfx)\;,
  \end{aligned}\qquad \Bx\in\Omega_{s},\;\rfx := \Phibf_{s}^{-1}(\Bx)\;.
\end{gather}
Next, we rewrite the more explicit force formula  \cref{eq:37}+\cref{eq:36}+\cref{eq:38}
in the ``classical fashion''
\begin{multline}
  \label{eq:33m}
  \int\nolimits_{\Omega}\Vf(\Bx)\cdot\velo(\Bx)\,\mathrm{d}\Bx \\ =
  \begin{aligned}[t]
    & \int\nolimits_{\partial\Omega}\fen(\Bx,\VB(\Bx),r(\Bx))(\velo(\Bx)\cdot\Bn(\Bx))\,\mathrm{d}S(\Bx)
    - \\ & \int\nolimits_{\Omega}
    \begin{aligned}[t]
      \Bigl\{ & \grad_{\Bx}\fen(\Bx,\VB(\Bx)),r(\Bx))\cdot \velo(\Bx)  -       \\ 
      & \grad_{\Vb}\fen(\Bx,\VB(\Bx)),r(\Bx))\cdot \curl(\VB\times \velo)(\Bx) + \\
      & \frac{\partial \fen}{\partial r}(\Bx,\VB(\Bx)),r(\Bx))\,\Div(r\,\velo)(\Bx)\Bigr\}
      \,\mathrm{d}\Bx -
    \end{aligned} \\
    & \int\nolimits_{\partial\Omega}(\VH(\Bx,\VB(\Bx),r(\Bx))\cdot \curl \VP(\Bx))
    (\velo(\Bx)\cdot\Bn(\Bx))\,\mathrm{d}S(\Bx)
    - \\ & \int\nolimits_{\Omega}
    \begin{aligned}[t]
      \Bigl\{\Bigl(& \grad_{\Bx}\VH(\Bx,\VB(\Bx),r(\Bx))\cdot\velo(\Bx) + \\ &
      \grad_{\Vb}\VH(\Bx,\VB(\Bx),r(\Bx))\cdot \curl(\VB\times \velo)(\Bx) + \\ &
      \frac{\partial\VH}{\partial r}(\Bx,\VB(\Bx),r(\Bx))\, \Div(r\,\velo)(\Bx)\Bigr)\cdot
      \curl \VP(\Bx)  + \\ &
      \VH(\Bx,\VB(\Bx),r(\Bx))\cdot \curl(\VB\times\velo)(\Bx)\Bigr\}\,\mathrm{d}\Bx
    \end{aligned}
  \end{aligned}
\end{multline}

\section{Special Case: Linear Materials}
\label{sec:lin}

The articles referring to passively advected ``frozen'' fields that we cited in the
Introduction all fail to mention an adjoint solution, which is crucial ingredient of
\cref{eq:main}. In this section we will explain, why this is not surprising in the case of
linear material laws. 

\subsection{Quadratic field energy}
\label{ss:wenl}

We restrict ourselves to local linear material laws merely depending on the spatial
coordinate $\Bs$. In this simplest case the local field energy density
$\fen=\fen(\Bx,\omega,\rho)$ as introduced in \cref{eq:fen} is a homogeneous quadratic
form in its second argument. Concretely, we assume
\begin{gather}
  \label{eq:24}
  \fen(\Bx,\omega,\rho) = \tfrac12 \Aloc(\Bx;\omega,\omega)\;,\quad
  \Bx\in \Omega,\;\omega\in \Lambda^{\ell+1}(\bbR^{3})
  \;, 
\end{gather}
with
$\Aloc:\Omega\times \Lambda^{\ell+1}(\bbR^{3})\times \Lambda^{\ell+1}(\bbR^{3})\to
\Lambda^{3}(\bbR)$, so that $\hop\Aloc$ is a symmetric $\Bx$-uniformly positive definite
\emph{bilinear form} in the last two arguments. Then, specializing \cref{eq:wen}, the
total field energy is
\begin{gather}
  \label{eq:25}
  \wenl(\varphibf) := \tfrac12 \int\nolimits_{\Omega}
  \Aloc(\Bx,(\extd\varphibf)(\Bx),(\extd\varphibf)(\Bx)) \;,
\end{gather}
where $\varphibf\in \Hdf$ solves the particular linear state variational equation
\begin{gather}
  \label{eq:26}
  \int\nolimits_{\Omega}\Aloc(\Bx,(\extd\varphibf)(\Bx),(\extd\etabf)(\Bx))
  = \int\nolimits_{\Omega}\sigmabf\wedge\etabf\quad\forall \etabf\in \Hdf \;.
\end{gather}
This setting allows considerable simplifications of the formulas derived in
\Cref{ss:sdfe}.

\begin{remark}
  \label{rem:linml}
  The local material law $\lml$ according to \cref{eq:lml} is obviously linear in the
  field argument,
  \begin{gather}
    \label{eq:27}
    \lml(\Bx,\omega)\wedge\nu = \Aloc(\Bx,\omega,\nu) \quad\forall \nu\in
    \Lambda^{\ell+1}(\bbR^{3}) \;.
  \end{gather}
\end{remark}

\subsection{Adapted adjoint method}
\label{ss:asma}

The simplifications due to the linear setting are most conspicuous in the abstract
framework of \Cref{ss:adjm}.  Adopting the notations of that section we now present the
adjoint approach for differentiation under variational constraints for the special case of
the functional
\begin{gather}
  \label{eq:14}
  F:\bbR\times H \to \bbR\;,\quad F(s,\Vv) := \tfrac12 \blf{a}(s;\Vv,\Vv) \;,
\end{gather}
where ${\blf{a}:\bbR\times H \times H \to \bbR}$ is a bounded, symmetric, $s$-uniformly $H$-elliptic
bilinear form in its last two arguments, and continuously differentiable in $s$. Then the
state variational problem \cref{eq:6} becomes
\begin{gather}
  \label{eq:18}
  \Vu(s)\in H:\quad
  \blf{a}(s;\Vu(s),\Vv) = \left<\ell(s),\Vv\right>\quad\forall \Vv\in H \;.
\end{gather}
Since
\begin{align}
  \label{eq:21}
  \left<
    \frac{\partial F}{\partial \Vw}(s,\Vw),\Vq
  \right> & = \blf{a}(s;\Vw,\Vq) \;,\\
  \label{eq:23}
   \left<
    \frac{\partial^{2}F}{\partial\Vw^{2}}(s,\Vw),(\Vv,\Vq)
  \right> & = \blf{a}(s;\Vv,\Vq) \;,
\end{align}
we obtain as specialization of the adjoint variational probem \cref{eq:12}
\begin{gather}
  \label{eq:20}
  \Vp\in H:\quad \blf{a}(0;\Vp,\Vq) = - \blf{a}(0;\Vu(0),\Vq)\quad\forall \Vq\in H \;,
\end{gather}
with the obvious adjoint solution $\boxed{\Vp=-\Vu(0)}$. Next, using \cref{eq:14} and
\cref{eq:21}, Formula \cref{eq:12} simplifies to
\begin{align}
  \notag
  \frac{d\wt{F}}{ds}(0) & = \frac{d}{ds}
  \left\{
    \tfrac12 \blf{a}(s;\Vu(0),\Vu(0)) - \blf{a}(s;\Vu(0),\Vu(0)) + \left<\ell(s),\Vu(0)\right>
  \right\}_{|s=0}  \\
  \label{eq:22}
  &  = \frac{d}{ds} \left\{-\tfrac12
    \blf{a}(s;\Vu(0),\Vu(0)) +  \left<\ell(s),\Vu(0)\right> \right\}_{s=0} \;.
\end{align}

\subsection{Force formulas}
\label{ss:lf}

As in \Cref{sec:vwm} we pursue the velocity method of shape calculus and employ pullback
to the ``reference domain'' $\Omega$ as in \Cref{ss:trf}, which yields, in analogy to
\cref{eq:5}, the total field energy for configurations deformed under the flow map induced
by $\velo$:
\newcommand{\Aloct}{\wt{\Ca}_{\mathrm{loc}}}
\begin{gather}
  \label{eq:Wtrfl}
  \wh{\wen}_{L}(s,\wh{\varphibf}) = \tfrac12
  \int\nolimits_{\Omega} \Phibf_{s}^{*}\left\{
    \Bx\mapsto \Aloct(s)(\Bx,(\extd(\Phibf_{s}^{-1})^{*}\wh{\varphibf}(s))(\Bx),
    (\extd(\Phibf_{s}^{-1})^{*}\wh{\varphibf}(s))(\Bx))\right\}
    \;,
\end{gather}
where $\Aloct(s)(\Bx,\omega,\omega) := \Aloc(\Phibf_{s}^{-1}(\Bx),\omega,\omega)$,
$\Bx\in\Omega_{s}$, and $\wh{\varphibf}(s)\in\Hdf$ is the solution of a transformed state
variational problem corresponding to \eqref{eq:vfft}:
\begin{multline}
  \label{eq:vfftl}
  \int\nolimits_{\Omega} \Phibf_{s}^{*}\left\{\Bx\mapsto \Aloct(s)(\Bx,
  \left(\extd(\Phibf_{s}^{-1})^{*}\wh{\varphibf}(s)\right)(\Bx),
  \left(\extd(\Phibf_{s}^{-1})^{*}{\etabf}\right)(\Bx))\right\}\\
   = \int\nolimits_{\Omega}\sigmabf\wedge{\etabf}\quad\forall {\etabf}\in \Hdf \;.
\end{multline}
Connecting these formulas with those of \Cref{ss:asma} we find that we can apply the
abstract theory of \Cref{ss:asma} when substituting $H \leftrightarrow \Hdf$,
\begin{align}
  \label{eq:abd}
  \blf{a}(s,\Vw,\Vv)\quad \leftrightarrow\quad & (s,\nubf,\etabf) \mapsto 
  \int\nolimits_{\Omega}
  \begin{aligned}[t]
     \Phibf_{s}^{*}\Bigl\{\Bx\mapsto \Aloct(s)(&\Bx,(\extd(\Phibf_{s}^{-1})^{*}{\nubf})(\Bx),\\
    & (\extd(\Phibf_{s}^{-1})^{*}{\etabf})(\Bx))\Bigr\}\;,
  \end{aligned}\\
  \label{eq:29}
  \left<\ell(s),\Vw\right>\quad \leftrightarrow\quad &
  \etabf \mapsto \int\nolimits_{\Omega}\sigmabf\wedge{\etabf} \;.
\end{align}
Hence, \cref{eq:22} gives us the force 1-form
\begin{align*}
  \left<\phibf,\velo\right>
  & =  - \tfrac12 \frac{d}{ds}\Biggl\{\int\nolimits_{\Omega}
  \begin{aligned}[t]
    \Phibf_{s}^{*}\Bigl\{\Bx\mapsto \Aloct(s)(\Bx,
    (\extd(\Phibf_{s}^{-1})^{*}\wh{\varphibf})(\Bx),
    (\extd(\Phibf_{s}^{-1})^{*}\wh{\varphibf})(\Bx))\Bigr\}\Biggr\}_{|s=0} \;,
  \end{aligned}
\end{align*}
where $\wh{\varphibf}$ is the solution of the state variational equation \cref{eq:26}.
Relying on the \emph{frozen} solution of the state variational problem
${\froz{\varphibf}(s) := (\Phibf_{s}^{-1})^{*}\wh{\varphibf}}$, defined analogously to
\cref{eq:13}, as in \Cref{ss:sdfe} we can equivalently write the force formula as
\begin{multline}
  \label{eq:28}
  \boxed{\left<\phibf,\velo\right>  
  =  -\tfrac12\frac{d}{ds}\Bigl\{
  \int\nolimits_{\Omega_{s}}\Aloc(\Phibf_{s}^{-1}(\Bx),
  (\extd \froz{\varphibf}(s))(\Bx),
  (\extd \froz{\varphibf}(s))(\Bx))
  \Bigr\}_{|s=0} 
  }\;.
\end{multline}
This formula expresses that
\medskip

\begin{center}
  \mycolorbox{.9\linewidth}{
    the force ``in direction $\velo$'' is the $s$-derivative in $s=0$ of the (negative) energy
    of the field solution passively advected in the $\velo$-flow.
  }
\end{center}
\medskip

As in \Cref{ss:sdfe} and under the assumption that $\varphibf$ enjoys sufficient
regularity, the chain rule, \cref{eq:16} and \cref{eq:Lie} allow to rewrite \cref{eq:28}
into
\begin{align}
  \notag
  \left<\phibf,\velo\right> & = -
  \begin{aligned}[t]
    &  \tfrac12 \int\nolimits_{\partial\Omega} \contr \left\{\Bx\mapsto \Aloc(\Bx,
      (\extd {\varphibf})(\Bx),
      (\extd {\varphibf})(\Bx))\right\} - \\ & 
  \tfrac12 \int\nolimits_{\Omega} \frac{d}{ds} \left\{
    \Aloc(\Phibf_{s}^{-1}(\Bx),
  (\extd \froz{\varphibf}(s))(\Bx),
  (\extd \froz{\varphibf}(s))(\Bx))
  \right\}_{|s=0}
\end{aligned} \\
\label{eq:gff}
& = -
\begin{aligned}[t]
  &  \tfrac12 \int\nolimits_{\partial\Omega} \contr \left\{\Bx\mapsto\Aloc(\Bx,
   (\extd {\varphibf})(\Bx),
   (\extd {\varphibf})(\Bx))\right\}+ \\ &
   \tfrac12 \int\nolimits_{\Omega}
   \left<\frac{\partial\Aloc}{\partial
       \Bx}(\Bx,(\extd\varphibf)(\Bx),(\extd\varphibf)(\Bx)),\velo(\Bx)\right> + \\ &
   \phantom{\tfrac12}\int\nolimits_{\Omega} \Aloc(\Bx,(\extd\varphibf)(\Bx),(\Lie
   \extd\varphibf)(\Bx)) \;.
\end{aligned}
\end{align}
The Lie derivative commutes with the exterior derivative, $\Lie\extd = \extd\Lie$, which,
if $\Lie\varphibf \in \Hdf$, permits us to recast the last term using the variational
equation \cref{eq:26},
\begin{gather}
  \label{eq:42}
   \int\nolimits_{\Omega} \Aloc(\Bx,(\extd\varphibf)(\Bx),(\Lie
   \extd\varphibf)(\Bx)) = 
   \int\nolimits_{\Omega} \sigmabf\wedge \contr(\extd\varphibf) \;.
\end{gather}

\subsection{Special cases and vector proxy incarnations}
\label{sec:lsp}

We restrict ourselves to deformation fields that are compactly supported in $\Omega$,
$\velo\in C^{\infty}_{0}(\Omega)$, which makes the boundary terms in \cref{eq:gff} vanish.

\subsubsection{Homogeneous material} We assume that $\Aloc$ does not vary in space,
$\frac{\partial\Aloc}{\partial\Bx}=0$. Then, thanks to \cref{eq:42},
\begin{gather}
  \label{eq:43}
  \left<\phibf,\velo\right> = \int\nolimits_{\Omega} \sigmabf\wedge \contr (\extd\varphibf) \;. 
\end{gather}
Along the lines of \cref{ss:vpff}, now we want to express \cref{eq:43} in classical vector
analytic notation.

In the case of electrostatics, $\ell=0$, using \cref{eq:30} and the vector proxies of
\cref{eq:31}, we find
\begin{gather}
  \label{eq:46}
  \int\nolimits_{\Omega}\cop{\Vf}\cdot\velo\,\mathrm{d}\Bx = -
  \int\nolimits_{\Omega} (\cop{\rho_{i}\VE})\cdot\velo\,\mathrm{d}\Bx \;. 
\end{gather}
For magnetostatics, $\ell=1$, with the vector proxies of \cref{eq:32}, we obtain
\cite[Section~5]{BOS92}, \cite[Section~V]{BOS92f},
\begin{gather}
  \label{eq:47}
  \int\nolimits_{\Omega}\cop{\Vf}\cdot\velo\,\mathrm{d}\Bx = 
  \int\nolimits_{\Omega} \Vj_{i}\cdot(\velo\times \VB)\,\mathrm{d}\Bx  =
  - \int\nolimits_{\Omega} (\cop{\Vj_{i}\times \VB})\cdot\velo\,\mathrm{d}\Bx  
  \;. 
\end{gather}
We have recovered the formulas for the Coulomb force density and Lorenz force density!

\subsubsection{Material interface} Let $D$ be a Lipschitz domain strictly contained in
$\Omega$ with exterior unit normal vector field $\Bn$. We consider piecewise homogeneous
material,
\begin{gather}
  \label{eq:48}
  \Aloc(\Bx,\omega,\eta) = \alpha(\Bx)\left(\omega \wedge \hop \eta\right)\;,\quad
  \alpha(\Bx) = 
  \begin{cases}
    \alpha^{-} & \text{for }\Bx\in D \;,\\
    \alpha^{+} & \text{for }\Bx\in \Omega\setminus D \;,
  \end{cases}
\end{gather}
with $ \alpha^{-}, \alpha^{+} > 0$. In addition, we exclude deformation of the sources,
$\supp(\sigmabf)\cap \supp(\velo) = \emptyset$. This means that only the second term in
\cref{eq:gff} remains and it can be treated based on the considerations of \cref{rem:jump}
($\left[\cdot\right]_{\partial D}$ stands for the jump across $\partial D$):
\begin{multline}
  \label{eq:49}
  \left<\phibf,\velo\right> = \tfrac12 \int\nolimits_{\Omega}
  \left<\frac{\partial\Aloc}{\partial
      \Bx}(\Bx,(\extd\varphibf)(\Bx),(\extd\varphibf)(\Bx)),\velo(\Bx)\right>
  \\
  = \tfrac12 \int\nolimits_{\partial D} \left[\alpha\,\hop \left(
      (\extd\varphibf)(\Bx)\wedge\hop (\extd\varphibf)(\Bx)\right)\right]_{\partial D}
  \,\left(\velo(\Bx)\cdot\Bn(\Bx)\right)\,\mathrm{d}S(\Bx) \;.
\end{multline}
The force field turns out to be a distribution supported on $\partial\Delta$ and
``pointing in normal direction''. 

The Euclidean Hodge operator is an identity operation of (Euclidean) vector proxies,
whereas the $\wedge$-product becomes the Euclidean inner product.

Thus, in the case of elctrostatics, $\ell=0$, again invoking the vector proxies from
\cref{eq:31}, \cref{eq:49} turns out as
\begin{gather}
  \label{eq:50}
  \int\nolimits_{\Omega}\cop{\Vf}\cdot\velo\,\mathrm{d}\Bx =
  \tfrac12 \int\nolimits_{\partial D}\cop{\left[\epsilon \N{\VE}^{2}\right]_{\partial D}(\Bx)\,
    \Bn(\Bx)}\cdot\velo(\Bx)\,\mathrm{d}S(\Bx)\;.
\end{gather}
Here we have introduced $\epsilon$ as a replacement for $\alpha$, in order to comply with
notational conventions.  We see that the force field is concentrated on $\partial D$.

In the case of magnetostatics, $\ell=1$, in terms of the vector proxies from \cref{eq:32},
the formula is, \cite[Formula~(12)]{BOS11},
\begin{gather}
  \label{eq:50m}
  \int\nolimits_{\Omega}\cop{\Vf}\cdot\velo\,\mathrm{d}\Bx =
  \tfrac12 \int\nolimits_{\partial D}\cop{\left[\chi\N{\VB}^{2}\right]_{\partial
      D}(\Bx)\,\Bn(\Bx)}
  \cdot\velo(\Bx)\,\mathrm{d}S(\Bx)\;.
\end{gather}

\section{Conclusion} 
\label{sec:concl}

We took for granted that for stationary electromagnetic field problems the virtual
work principle based on field energies offers a general way to define and compute
mechanical forces effected by electromagnetic fields. We gave a rigorous mathematical
treatment of that principle using techniques from shape calculus and differentiation under
variational constraints by means of the adjoint method.

The main result is that the force field viewed as a vector-valued distribution can be
computed by shape differentiating appropriate field energy functionals with \emph{merely
passively advected electromagnetics fields} (``Holding the fields constant''). In the
case of non-linear materials some of these fields have to be computed as solutions of
adjoint variational problems.

Only in the case of linear materials all fields are ``physical'' in the sense that they
solve the state variational problem, that is, the weak formulation of the static field
equations. Hence,\\
\mycolorbox{0.975\linewidth}{only in the linear case the customary recipe of passively advecting the
fields, amounts to a correct application of the virtual work principle. For non-linear
materials skipping the adjoint problem will lead to wrong solutions.}


\end{document}